\documentclass[12pt,a4paper]{article}
\pdfoutput=1

\usepackage[utf8]{inputenc}

\usepackage{multirow}
\usepackage{amsmath}
\usepackage{color}
\usepackage{amsfonts}
\usepackage{amssymb}
\usepackage{graphicx}
\usepackage{geometry}
\usepackage{amssymb,epsfig,subfigure}
\usepackage{hyperref}
\usepackage{url}
\usepackage{comment}
\usepackage[font=footnotesize]{caption}
\usepackage{slashed}
\usepackage{amssymb}
\usepackage{amsfonts}
\usepackage{bbm}
\usepackage{bm}

\usepackage{booktabs}

\usepackage{cite}

\usepackage{calc}

\usepackage{accents}
\newcommand{\dbtilde}[1]{\accentset{\approx}{#1}}

\makeatletter
\renewcommand\section{\@startsection {section}{1}{\z@}%
                                 {-3.5ex \@plus -1ex \@minus -.2ex}%nn
                                   {2.3ex \@plus.2ex}%
                                   {\normalfont\large\bfseries}}
\renewcommand\subsection{\@startsection{subsection}{2}{\z@}%
                                   {-3.25ex\@plus -1ex \@minus -.2ex}%
                                     {1.5ex \@plus .2ex}%
                                     {\normalfont\bfseries}}
\renewcommand\subsubsection{\@startsection{subsubsection}{3}{\z@}%
                                   {-3.25ex\@plus -1ex \@minus -.2ex}%
                                     {1.5ex \@plus .2ex}%
                                     {\normalfont\itshape}}
\makeatother

\def\pplogo{\vbox{\kern-\headheight\kern -29pt
\halign{##&##\hfil\cr&{\ppnumber}\cr\rule{0pt}{2.5ex}&\ppdate\cr}}}
\makeatletter
\def\ps@firstpage{\ps@empty \def\@oddhead{\hss\pplogo}%
  \let\@evenhead\@oddhead % in case an article starts on a left-hand page
}%      The only change in \maketitle is \thispagestyle{firstpage} instead of 
\def\maketitle{\par
 \begingroup
 \def\thefootnote{\fnsymbol{footnote}}
 \def\@makefnmark{\hbox{$^{\@thefnmark}$\hss}}
 \if@twocolumn
 \twocolumn[\@maketitle]
 \else \newpage
 \global\@topnum\z@ \@maketitle \fi\thispagestyle{firstpage}\@thanks
 \endgroup
 \setcounter{footnote}{0}
 \let\maketitle\relax
 \let\@maketitle\relax
 \gdef\@thanks{}\gdef\@author{}\gdef\@title{}\let\thanks\relax}
\makeatother

\usepackage{titlesec}

\titleformat{\subsubsection}
  {\normalfont\normalsize\bfseries}
  {\thesubsubsection}
  {1em}
  {}

\numberwithin{equation}{section}

\newcommand\eea{\end{eqnarray}}
\newcommand\bea{\begin{eqnarray}}

\def\beq{\begin{equation}}
\def\eeq{\end{equation}}

\newcommand{\be}{\begin{equation}}
\newcommand{\ee}{\end{equation}}
\newcommand{\ba}{\begin{align}}
\newcommand{\ea}{\end{align}}
\newcommand{\bg}{\begin{gather}}
\newcommand{\eg}{\end{gather}}
\newcommand{\bseq}{\begin{subequations}}
\newcommand{\eseq}{\end{subequations}}

\newcommand{\Tr}{{\rm Tr}}
\renewcommand{\Im}{\mathop{\rm Im}\nolimits}
\renewcommand{\Re}{\mathop{\rm Re}\nolimits}

\newcommand{\tr}{{\rm tr}}
\newcommand{\mc}{\mathcal}

\newcommand{\dd}[1]{\text{d}#1}

\newcommand{\ket}[1]{\left|#1\right\rangle}

\newcommand{\ketbra}[2]{\left|#1\right\rangle\!\!\left\langle#2\right|}

\newcommand{\expval}[3]{\left\langle#1\right|#2\left|#3\right\rangle}
\newcommand{\of}[1]{\left(#1\right)}
\newcommand{\off}[1]{\left[#1\right]}
\renewcommand{\exp}[1]{\,\mathrm{exp}\of{#1}}

\newcommand{\coment}[1]{}

\begin{document}
\setcounter{page}0
\def\ppnumber{\vbox{\baselineskip14pt
%\hbox{hep-th/0000000}
}}
\def\ppdate{
%\footnotesize{SU/ITP-14/XX}
} \date{}

\author{Nicolás Abate$^{1, 2 }$ and Leandro Martinek$^{1, 2}$\\
[7mm] \\
{\normalsize \it $^1$Centro At\'omico Bariloche and CONICET}\\
{\normalsize \it $^2$ Instituto Balseiro, UNCuyo and CNEA}\\
{\normalsize \it S.C. de Bariloche, R\'io Negro, R8402AGP, Argentina}\\
}

\bigskip
\title{\bf  Exact mutual information for free\\ massless fermions in any dimension
\vskip 0.5cm}
\maketitle

\begin{abstract}
We obtain an exact expression for the Rényi mutual information (RMI) between two spherical regions for a free, massless Dirac field in $d$ spacetime dimensions and for a massless Rarita-Schwinger field in $d=4$. We identify each angular momentum mode in the spherical dimensional reduction of the parent theory with a massive Dirac fermion in two dimensional anti-de Sitter space (AdS). The RMI for each mode can be computed analytically for any Rényi index $n$ in terms of solutions of a Painlevé differential equation. We explore the long and short distance asymptotics of the result, where the sum over modes can be computed explicitly. As a result we are able to find subleading contributions to the long distance expansion in $d$ dimensions for arbitrary $n$, as well as to extract the area and logarithmic coefficients for the case $n=1$ in the short distance regime.
\end{abstract}

\newpage

\tableofcontents

\begin{center}
    \rule{0.8\textwidth}{0.4pt}
\end{center}

\section{Introduction}

The vacuum mutual information is a fundamental quantity of information theory measuring both quantum and classical correlations among two subsystems. A usual way of computing this quantity starts with the Rényi entropy of some subsystem $\mc V$, namely
\begin{equation}
    S_{n}(\mc V) = (1-n)^{-1} \log \tr \,\rho_{\mc V}^{n}\,,
\label{eq:nRenyiEntropy}
\end{equation}
where $\rho_\mc V$ is the reduced density matrix associated to $\mc V$ and the vacuum state. Then, the Rényi mutual information (RMI) between two subsystems $\mc A$ and $\mc B$ is introduced as
\begin{equation}
    I_{n}(\mc A,\mc B) = S_{n}(\mc A) + S_{n}(\mc B) - S_{n}(\mc A \cup \mc B) \,.
\label{eq:nRenyiMutual}
\end{equation}
The Rényi entropy encodes information about the spectrum of the reduced state and is easily calculable using the so--called replica trick, as we review below. In the limit $n\to1$, it gives the entanglement entropy $S(\mc V)$ between the subsystem $\mc V$ and its complement. Correspondingly, the RMI in the $n\to1$ limit gives the mutual information
\begin{equation}
    \lim_{n \rightarrow 1} I_{n}(\mc A,\mc B) \equiv I(\mc A,\mc B) = S(\mc A) + S(\mc B) - S(\mc A \cup \mc B) \, .
\end{equation}

In the context of Quantum Field Theory (QFT), the subsystems are regions of spacetime. It has been argued that the mutual information has the potential to completely characterize the underlying theory. Namely, when the regions are close to each other the mutual information serves as a geometrical regulator for entanglement entropy and thus captures all its cutoff independent contributions, such as the renormalization group charges \cite{Casini:2008wt,Casini:2015woa,Casini:2023kyj}. When the regions are distant spheres and when the model is a conformal field theory (CFT), it admits an expansion as a power series whose exponents and coefficients encode the scaling dimension of primary operators and the operator product expansion (OPE) data, respectively \cite{Calabrese:2010he,Cardy:2013nua,Agon:2015twa,Long:2016vkg,Chen:2016mya,Chen:2017hbk}. Therefore, lately a lot of effort has been made to both compute mutual informations for simple theories, and also to understand how to reconstruct the underlying theory from the knowledge of the mutual information between two arbitrary regions \cite{Agon:2021zvp,Agon:2024xvs,Agon:2024zae}.

\begin{figure}[h]
    \centering
    \includegraphics[width=0.6\linewidth]{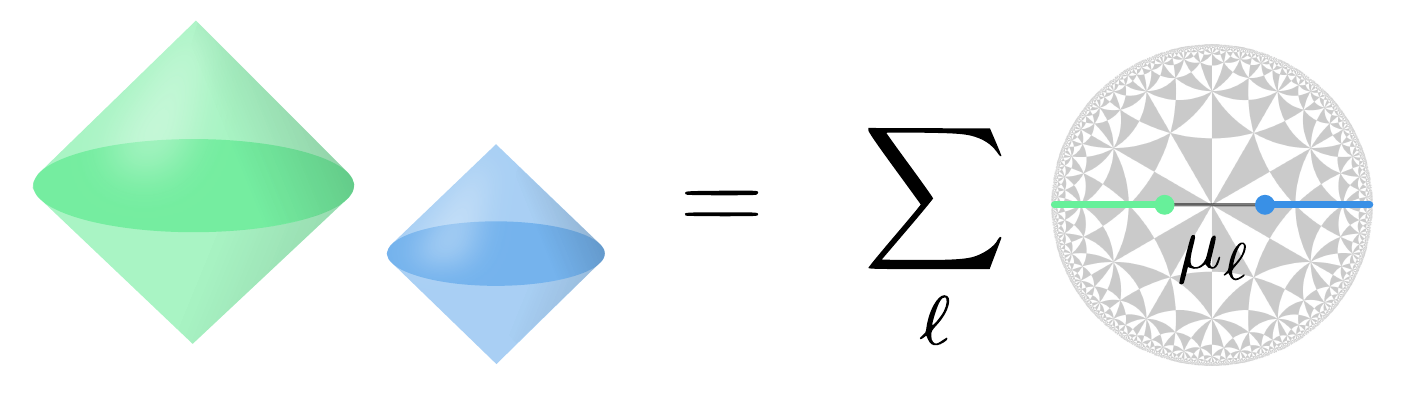}
    \caption{The main idea of our work consists in describing the mutual information between two spherical regions for the massless theory in $d$ dimensions as a sum of mutual informations for massive theories in AdS$_2$.}
    \label{fig:intro}
\end{figure}

In this work we develop a strategy for exactly computing the RMI between spherical regions for free, massless fermionic theories in any spacetime dimension. The main idea is illustrated in Fig. \ref{fig:intro} and goes as follows. We start by performing a spherical dimensional reduction of the theory which transforms the massless problem in $d$ dimensions into an infinite tower of massive theories living on the half line $r \in (0, \infty)$. This gives the following expression for the RMI
\be\label{eq:MIsumintro}
I_n(\mc A,\mc B)=\sum_{\ell} \#_\ell I_n(A,B;\mu_\ell)\,,
\ee
where $A$ and $B$ are the radial projections of the spherical regions and $\mu_\ell$ is related to the mass of the theory on the half line, labeled by the angular momentum value $\ell$. 
This multipolar expansion is a standard procedure in the literature, dating back to the seminal work by Sredniki \cite{Srednicki:1993im}, that allows for numerical computations in the lattice. We give a step forward in this direction, finding an analytic expression for the RMI between $A$ and $B$ for each one of the theories on the half line. We show that they are equivalent to a Dirac fermion propagating in a two dimensional Anti-de Sitter (AdS$_2$) spacetime, where it is possible to use results by Palmer et al. and Doyon about the two point function of twist operators\cite{Palmer:1993jp,Doyon:2003nb}. Therefore, we are able to compute the sum \eqref{eq:MIsumintro} to obtain the RMI for the theory in $d$ dimensions.

The richness of Anti-de Sitter spacetime also helped understanding the physics occurring in flat space in many other, completely different contexts. A remarkable example is of course the AdS/CFT correspondence \cite{Duetsch:2002hc,Ryu:2006bv,Faulkner:2013ana}. However, as we do in this work, one does not need to restrict to holography. The study of QFT in a fixed AdS background has become increasingly popular lately, taking advantage of the large amount of symmetries the spacetime possesses albeit having negative curvature. Modern bootstrap approaches formulated in AdS allow to obtain S-matrix results in the flat space limit \cite{Paulos:2016fap,Komatsu:2020sag,vanRees:2022zmr}, the effects of adding boundaries and defects to conformal theories becomes transparent in AdS \cite{Bason:2025sxb,Giombi:2020rmc,Carmi:2018qzm} and also confinement in AdS is being actively studied \cite{Copetti:2023sya,Ciccone:2024guw,Ciccone:2025dqx}. Regarding quantum information, these techniques have been applied to AdS only very recently \cite{Boutivas:2025ksp,Abate:2026apg}.

We think that the relevance of our results are twofold. On the one hand, we provide an exact result for a RMI in arbitrary spacetime dimensions, while exact results for quantum information measures are sparse in the literature. In \cite{Huerta:2022tpq,Huerta:2023dqt} an analytic expression for the entanglement entropy of the sphere was obtained as a sum of entropies of an interval attached to the origin on the half line for each angular momentum $\ell$, in the same spirit as \eqref{eq:MIsumintro}. Analytic expressions for the mutual information are only known for $1+1$ dimensional, free models \cite{Arias:2018tmw, Mintchev:2020uom, Blanco:2021kzm}, while for higher dimensional theories only asymptotical approximations \cite{Calabrese:2010he,Cardy:2013nua, Hertzberg:2010uv, Casini:2014yca}, large $N$ expansions \cite{Headrick:2010zt,Fischler:2012uv} or differences \cite{Abate:2025ywp} are known. On the other hand, our identification between the radial problem and the model in AdS$_2$ provides yet another example of the usefulness of understanding QFT in this background. 

This article is structured as follows. We describe the dimensional reduction that gives rise to the multipolar expansion for the RMI between two spheres in full detail in Section \ref{sec:dimReg}. We mostly focus on the spin 1/2 case in arbitrary dimension $d$, but we also comment on the spin 3/2 case for $d=4$. In Section \ref{sec:AdS} we give the exact result for each angular momentum contribution in the multipolar expansion by mapping the problem to that of computing RMIs in AdS$_2$. We begin by establishing the equivalence between the dimensional reduced theories and the two dimensional models in the curved background. Then, we derive expressions for the $n$th RMI in AdS$_2$, including the analytical continuation to $n\to1$, and also comment on the long and short distance limits of these results. Later, in section \ref{sec:d_dimensions}, we combine the results of the previous sections to write the exact RMI for the $d$ dimensional parent theory. We show that the sum over angular momentum becomes manageable when considering spheres very far apart or very close to each other. In the former we recover the expected power law behavior expeted for a $d$ dimensional CFT, and in the latter we obtain the universal area and logarithmic coefficients appearing in the mutual information. We conclude with a summary of our results in Section \ref{sec:conclusions} and comment on some future directions.

\section{Dimensional reduction for massless fermions}\label{sec:dimReg}

A general fact about free quantum field theories in $d$ dimensional flat spacetimes is that their Hamiltonians can be decomposed as a sum of two dimensional contributions, each associated to a radial problem for different angular momentum modes. In this Section, we will review this fact for free massless theories describing particles with half-integer spin. We will mostly focus on the Dirac field (with spin 1/2) which is a CFT in arbitrary dimensions, but we will also briefly comment on this procedure for the Rarita-Schiwinger field (spin 3/2) in $d=4$ where this model also acquires conformal invariance.

\subsection{Dirac field}

Consider first a massless Dirac field in $\mathbb{R}^{d-1,1}$, and let us work using spherical coordinates. Spinors in general coordinate systems can be described by introducing a veilbein $e^a=e^a_\mu\,\dd x^\mu$  (here the latin index $a$ ranges in $0,\dots,d-1$), defined by
\be 
g_{\mu\nu}=\eta_{ab}e_\mu^a e_\nu^b\,.
\ee
Using this, one can define the curved-space gamma matrices $\gamma^\mu$ in a locally inertial frame as $\gamma^\mu(x)=e^\mu_a(x)\gamma^a$, where $\gamma^a$ are their flat-spacetime counterparts. The vielbein also defines the spin connection $\omega_\mu^{ab}$ via the Cartan relation\footnote{Or equivalently $\omega_\mu^{ab}
=
e^{a\nu}\left(\partial_{\mu} e_{\nu}{}^{b}
-
\Gamma^{\lambda}_{\mu\nu} e_{\lambda}^{b}\right)$, where $\Gamma^\lambda_{\mu\nu}$ is the Levi-Civita connection.}
\be\label{eq:cartan} 
\dd e^a+\eta_{bc}\,\omega_\mu^{ac}\,\dd x^\mu=0\,.
\ee
Then, a covariant derivative $D_\mu$ is introduced. Its action on the spinor fields $\Psi$ and $\bar\Psi$ is
\be
\begin{aligned}
D_\mu\Psi&=\partial_\mu\Psi+\frac{1}{8}\omega_\mu^{ab}\,[\gamma^a,\gamma^b]\Psi\,,\\
D_\mu\bar\Psi&=\partial_\mu\bar\Psi-\frac{1}{8}\omega_\mu^{ab}\,\bar\Psi[\gamma^a,\gamma^b]\,.
\end{aligned}
\ee
These structures allow us to write the (symmetrized) action as
\be\label{eq:action_Dirac_gral}
\mc S=\frac{i}{2}\int\dd^dx\ \sqrt{|g|}\of{\bar\Psi\gamma^\mu D_\mu\Psi-D_\mu\bar\Psi\gamma^\mu\Psi}\,.
\ee

Following \cite{Huerta:2023dqt} and relying on the conformal invariance of the theory, we will work with the metric
\be\label{eq:misterious_metric} 
\dd s^2=\frac{\dd t^2-\dd r^2}{r^2}+\dd\Omega_{d-2}^2\,,
\ee
which is related to the usual metric for spherical coordinates via a Weyl transformation with conformal factor $r^{-2}$. The advantage in doing so is that in these coordinates, the spin connection and thus the covariant derivative acquire a block-diagonal form,\footnote{Note that the $t,r$ block of this metric matches the so called Poincaré patch of AdS. 
This coincidence will be important later on throughout our work.}
\be 
\omega_t^{01}=-\frac{1}{r}\,,\quad\omega_\Omega^{ab}=\of{1-\delta_0^a}(1-\delta_1^b)\,\omega_\Omega\,,
\ee
where $\omega_\Omega$ is the spin connection on the $d-2$ dimensional sphere $\mathbb{S}^{d-2}$. This means that, crucially, the components mixing the radial and angular parts vanish. We pick the following representation for the flat-space gamma matrices
\be
\begin{aligned}
\gamma^0 &= \sigma^{3} \otimes \mathbbm{1}_{2^{[(d-2)/2]}}, \\
\gamma^1 &= i\sigma^{2} \otimes \mathbbm{1}_{2^{[(d-2)/2]}}, \\
\gamma^2 &= i\sigma^{1} \otimes \hat{\gamma}^0, \\
&\ \,\vdots \\
\gamma^{d-1} &= i\sigma^{1} \otimes \hat{\gamma}^{\,d-3}\,,
\end{aligned}
\ee
written in terms of the gamma matrices of a $d-2$ dimensional flat euclidean space $\hat\gamma^a$. It turns out that then one can write
\be 
\slashed{D}=\slashed{D}_{(t,r)}\otimes \mathbbm{1}_{2^{[(d-2)/2]}}+i\sigma^1\otimes \slashed{D}_{\Omega}\,.
\ee
Here $\slashed{D}_{(t,r)}$ is the massless Dirac operator in the space with metric $r^{-2}(\mathrm{d}t^2-\mathrm{d}r^2)$ and $\slashed{D}_\Omega$ is that of $\mathbb{S}^{d-2}$. Let us introduce orthonormal eigenfunctions for the latter operator as
\be\label{eq:Z_lms}
\slashed{D}_\Omega\, Z_{\ell\vec{m}s}^{(\pm)}=\pm i\mu_\ell\, Z_{\ell\vec{m}s}^{(\pm)}\,,\qquad \mu_\ell=\ell+\frac{d-2}{2}\,,
\ee
for $\ell\in\mathbb{N}_0$. The $\pm$ account for the helicity of the field and the quantum numbers $\vec{m},s$ give the following degeneracy for each eigenvalue (see \cite{Camporesi:1995fb} for further details):
\be\label{eq:degeneracy} 
\lambda(\ell)=\frac{2^{[(d-2)/2]}(\ell+d-3)!}{\ell!(d-3)!}\,,
\ee
where the brackets denote the floor function. We note that this expression is a polynomial of degree $d-3$. We can use this basis to expand the Dirac field as
\be\label{eq:Dirac_expansion} 
\Psi(t,r,\Omega)=\frac{1}{r^{(d-2)/2}}\sum_{\ell\vec{m}s}\psi_{\ell\vec{m}s}^{(+)}(t,r)\otimes Z_{\ell\vec{m}s}^{(+)}(\Omega)+\psi_{\ell\vec{m}s}^{(-)}(t,r)\otimes Z_{\ell\vec{m}s}^{(-)}(\Omega)\,,
\ee
where $\psi_{\ell\vec{m}s}^{(\pm)}$ are two dimensional Dirac spinors. The prefactor $r^{-(d-2)/2}$ comes from the Weyl rescaling that turns the usual spherical coordinate metric to \eqref{eq:misterious_metric} and a factor that renders the anti-commutation relations cannonically normalized.

Then, plugging \eqref{eq:Dirac_expansion} into \eqref{eq:action_Dirac_gral} the action decouples as $\mc S=\sum_{\ell\vec{m}s}\mc S_{\ell\vec{m}s}^{(+)}+\mc S_{\ell\vec{m}s}^{(-)}$. Integrating the angular variables, each term on the sum is of the form
\be\label{eq:S_lms}
\mc S_{\ell\vec{m}s}^{(\pm)}=i\int\dd t\int_0^\infty\dd r\ \bar\psi_{\ell\vec{m}s}^{(\pm)}\off{\frac{1}{2}\of{\sigma^3\overset{\leftrightarrow}{\partial}_t+i\sigma^2\overset{\leftrightarrow}{\partial}_r}\mp \frac{\sigma^1\mu_\ell}{r}}\psi_{\ell\vec{m}s}^{(\pm)}\,.
\ee
Note that this action has acquired a position-dependent mass term, which also depends on the angular momentum and dimensionality of the parent theory. Curiously, it is also worth mentioning that this new term renders the actions \eqref{eq:S_lms} scale invariant. Ultimately, under the redefinition $\psi_{\ell\vec{m}s}^{(-)}\to\sigma^1\psi_{\ell\vec{m}s}^{(-)}$, the $\pm$ contributions in the action become identical and so the corresponding equations of motion. The solutions to these equations which remain regular at $r=0$ are superpositions of the following modes
\be\label{eq:solutions_radial} 
\psi_{\ell\vec{m}s}^{(\pm)}(t,r)=e^{-ikt}\sqrt{r}\,\binom{J_{\mu_\ell+1/2}(kr)}{-iJ_{\mu_\ell-1/2}(kr)}\,.
\ee
From here one can quantize these theories in the usual way by introducing creation and annihilation operators at a fixed time slice and thus setting a vacuum state $|\Omega\rangle_{\ell\vec ms}$.  This then defines all correlation functions. We emphasize that these vacua are just labeled by $\ell$, since there is no dependence on $\vec{m}$ or $s$ in \eqref{eq:solutions_radial}. In turn, the vacuum state of the $d$ dimensional theory will be given by the tensor product $|\Omega_d\rangle=\bigotimes_{\ell\vec ms}|\Omega\rangle_{\ell\vec m s}$, where all factors with the same value of $\ell$ are identical.

The Hamiltonian also decouples as $H=\sum_{\ell\vec{m}s}H_{\ell\vec{m}s}^{(+)}+H_{\ell\vec{m}s}^{(-)}$ with each term being
\be\label{eq:hamiltonian_dimReg_fermion} 
H_{\ell\vec{m}s}^{(\pm)}=i\int_0^\infty\dd r\off{{\psi_{\ell\vec{m}s}^{(\pm)}}
^\dagger\of{-\sigma^1\partial_r+\frac{\mu_\ell}{r}\,i\sigma^2}\psi_{\ell\vec{m}s}^{(\pm)}-\mathrm{h.c.}}.
\ee
This analysis tells us that the $d$ dimensional parent theory can be understood as an infinite tower of two dimensional models living on a half line with solutions remaining finite at its boundary, at the expense of the appearance of a position-dependent mass term. Note that these two dimensional theories appear completely decoupled in the action, hamiltonian and vacuum state, so by understanding them we could in principle completely characterize the parent ones. 

\begin{figure}[h]
    \centering
    \includegraphics[width=0.9\linewidth]{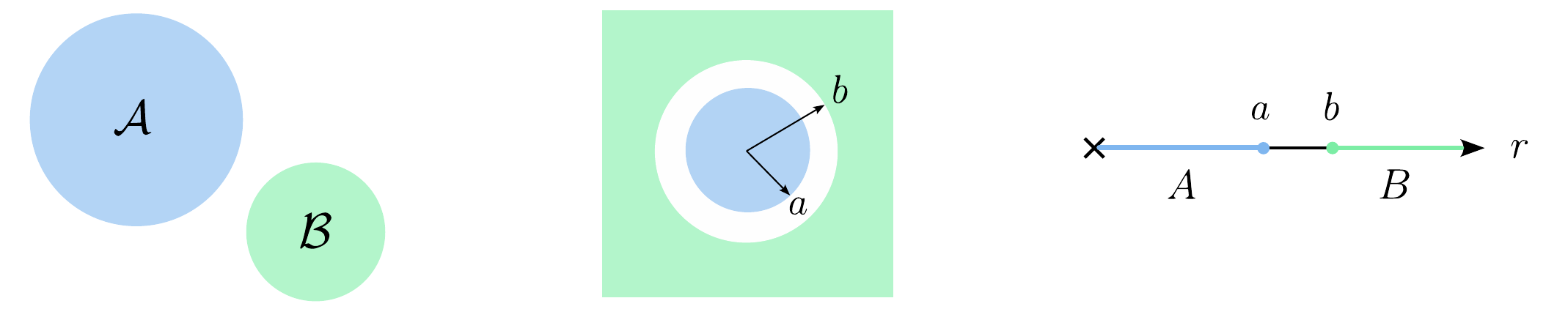}
    \caption{In the left panel we show two arbitrary spheres $\mc A$ and $\mc B$ in $d=4$ Minkowski spacetime. In the middle panel we show the image of $\mc A$ and $\mc B$ under the conformal transformation which maps them to the concentrical configuration preserving the cross ratio \eqref{eq:crossratio}. Finally on the right panel we show the dimensional reduction of the middle one, which consists of an arrangement of intervals on the half line.}
    \label{fig:conf}
\end{figure}

An important example of quantities of the $d$ dimensional theory that can be computed from the dimensionally reduced ones are Rényi entropies for regions with spherical symmetry and for the vacuum state. Namely
\be\label{eq:entropy_sum} 
S_n^{(d)}(\mc V)=\sum_{\ell=0}^\infty 2\lambda(\ell) S_n(V;\mu_\ell)\,,
\ee
where the factor 2 accounts for both helicity modes. Here $S_n^{(d)}(\mc V)$ represents the Rényi entropy of the Dirac fermion in $d$ dimensions for a spherically symmetric region $\mc V$ for the state $|\Omega_d\rangle$, while $S_n(V;\mu_\ell)$ is that of the two dimensional model with mass parameter $\mu_\ell$, computed for the radial projection of $\mc V$, which we denote as $V$ and will generically be a collection of intervals on the half line, and for the state $|\Omega\rangle_{\ell\vec ms}$. The Rényi entropies can then be used to compute RMIs for the Dirac fermion using \eqref{eq:nRenyiMutual}, which represents the ultimate goal of our work. More concretely, we will be interested in computing the RMI between two arbitrary spheres $\mc A$ and $\mc B$. Exploiting the conformal symmetry of the parent theory, we can map this configuration to that of a sphere of radius $a$ centered at the origin together with the outer part of another, concentrical sphere with radius $b>a$. In this transformation the conformal invariant is the cross-ratio
\begin{equation}\label{eq:crossratio}
    \eta = \frac{4ab}{(a+b)^{2}}\,, 
\end{equation}
and then $I_n^{(d)}(\mc A,\mc B)=I_n^{(d)}(\eta)$. Moreover, performing the dimensional reduction and using \eqref{eq:entropy_sum}, we can write
\be\label{eq:RMI_sum} 
I_n^{(d)}(\eta)=\sum_{\ell=0}^\infty2\lambda(\ell)I_n(A,B;\mu_\ell)\,,
\ee
where we have identified
\be\label{eq:RMI_half-line} 
I_n(A,B;\mu_\ell)=S_n((0,a);\mu_\ell)+S_n((b,\infty);\mu_\ell)-S_n((0,a)\cup(b,\infty);\mu_\ell)\,,
\ee
as the RMI for the two dimensional model with mass parameter $\mu_\ell$. Here $A=(0,a)$, $B=(b,\infty)$, see Fig. \ref{fig:conf} to clarify the geometrical setup.

\subsection{A comment on spin 3/2 fields}

Another interesting example is the Rarita-Schwinger model, where the dimensional reduction decomposes the Hamiltonian with the same structure as the Dirac case. The action for the Rarita-Schwinger massless theory is given by
\be 
\mc S=i\int\mathrm{d}^dx\ \Psi_\mu\gamma^{[\mu}\gamma^\nu\gamma^{\rho]}\partial_\nu\Psi_\rho\,,
\ee
in terms of a vector-spinor field $\Psi_\mu$, which is invariant under the gauge transformation $\Psi_\mu\to\Psi_\mu+\partial_\mu f$. This model describes particles with spin 3/2 and within supersymmetric contexts, it arises as the gravitino field. In $d=3$ dimensions one can show that the field has no gauge-invariant propagating degrees of freedom. In $d=4$ the theory becomes a CFT and thus Weyl invariance allows one to carry on an analogous procedure as in the previous subsection, while the lack of conformal invariance in $d\neq4$ makes this task impossible using our methods. 

The dimensional reduction of this model in $d=4$ was analyzed in detail in \cite{Benedetti:2023jye}, where the authors expanded the field in vector-spinor spherical harmonics and found that the Hamiltonian also decouples as two infinite towers of two dimensional spinorial modes as in \eqref{eq:hamiltonian_dimReg_fermion}. Explicitly,\footnote{In \cite{Benedetti:2023jye} a slightly different notation for the quantum numbers has been employed, however the multiplicity \eqref{eq:degeneracy} still accounts for every state in the decomposition.}
\be 
H=\sum_{\ell\vec{m}s}H_{\ell\vec{m}s}^{(+)}+H_{\ell\vec{m}s}^{(-)}\,,\quad H_{\ell\vec{m}s}^{(\pm)}=i\int_0^\infty\dd r\off{\psi_{\ell\vec{m}s}^{(\pm)}\of{-\sigma^1\partial_r+\frac{\mu_{\ell}}{r}\,i\sigma^2}\psi_{\ell\vec{m}s}^{(\pm)}-\mathrm{h.c.}},
\ee
where $\mu_\ell=\ell+1$ in $d=4$. The main difference with the spin 1/2 case is that the gauge invariance now implies that the $\ell=0$ mode is absent, so \eqref{eq:entropy_sum} and \eqref{eq:RMI_sum} get modified accordingly. Then the $\ell=0$ contribution gives the exact RMI difference between the Dirac and Rarita-Schwinger fields in $d=4$. This situation is similar to what happens between the Maxwell and scalar fields in their dimensional reduction \cite{Abate:2025ywp}, and again implies that studying the models on the half line is important to understand the higher dimensional case. 

\section{Free Dirac field in AdS$_2$}\label{sec:AdS}

For this section let us switch to a curved background. We will explicitly show that each term in the mode decomposition found in the previous section is actually equivalent to a free Dirac fermion propagating in AdS$_2$, with different masses for each angular momentum. Then we will focus on computing the Rényi mutual information for these models and derive some asymptotical limits.

\subsection{Geometry, boundary conditions and field equations}

\begin{figure}[h]
    \centering
    \includegraphics[width=0.67\linewidth]{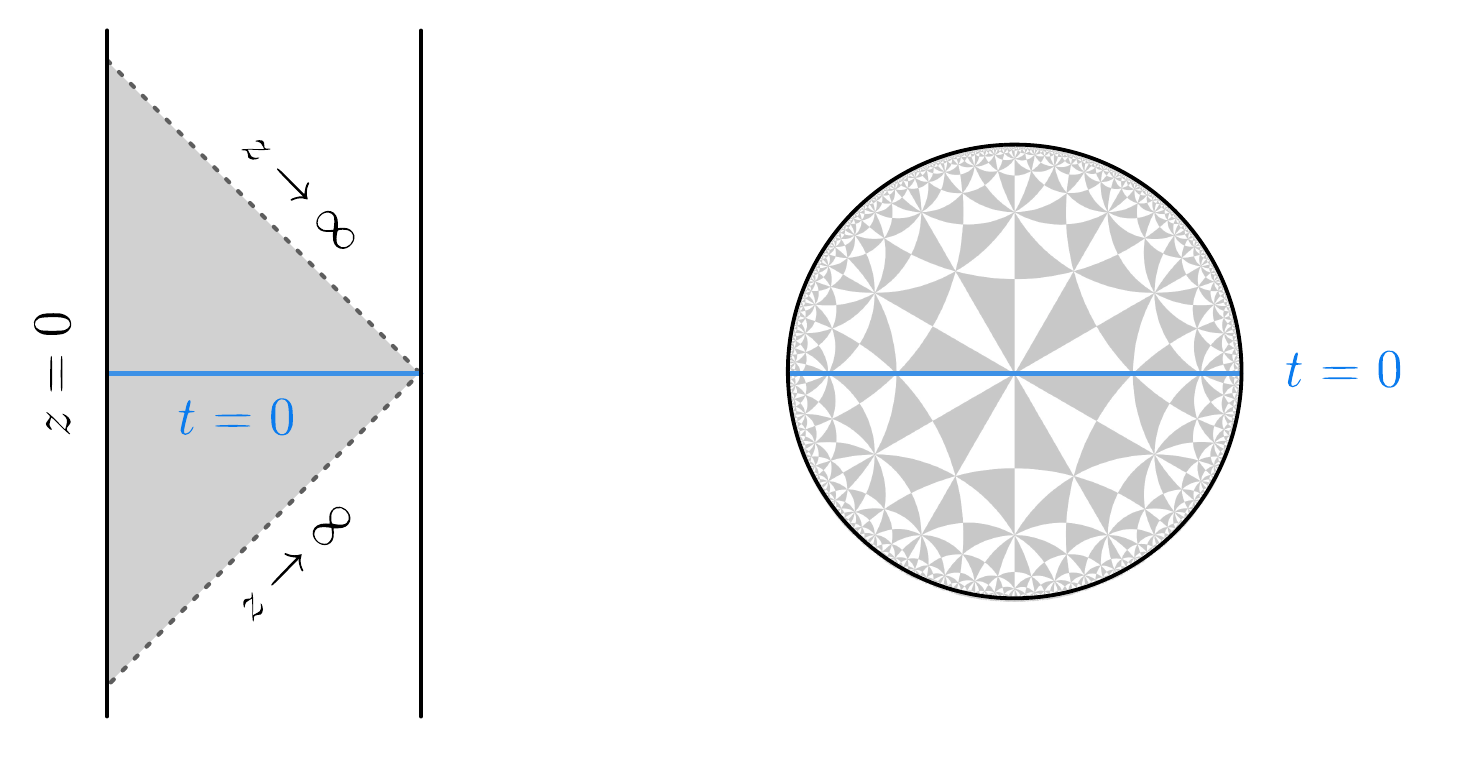}
    \caption{Penrose diagram for AdS$_2$ (left). The shaded region is the Poincaré Patch, described by the metric \eqref{eq:AdS2metric}. The Euclidean version of AdS$_2$ (right) is fully covered by the Poincaré Patch.}
    \label{fig:AdS2_penrose}
\end{figure}

Anti-de Sitter is a maximally symmetric spacetime. This means that it has the same amount of isometries as a Minkowksi spacetime with the same dimensionality. It also implies that the curvature is a constant, conventionally written in terms of some parameter with units of lenght $L$, which is called the AdS radius, as $\mc R=-d(d-1)L^{-2}$. In this work we will focus on the two dimensional case, and we will describe it using Poincaré coordinates
\be\label{eq:AdS2metric} 
\dd s^2=\frac{L^2}{z^2}\of{\dd t^2-\dd z^2},\quad z\geq0\,.
\ee
This coordinate system actually covers only a patch of the whole spacetime, as we show in the Penrose diagram of AdS$_2$ in Fig. \ref{fig:AdS2_penrose}. However, the euclidean version of AdS$_2$ is the hyperbolic space, and this is indeed fully coverd by the euclidean version of \eqref{eq:AdS2metric}. 

An important feature of AdS$_2$ is that it has two asymptotic timelike boundaries. As can be seen from \eqref{eq:AdS2metric}, one of the boundaries is located at $z=0$. Furthermore, note that the metric \eqref{eq:AdS2metric} is conformally equivalent to a Minkowski spacetime with boundary, thus CFTs in the Poincaré patch can be mapped to BCFTs in Minkowski. The symmetry group of the latter theories is $SO(2,1)$, which actually coincides with the isometry group of AdS$_2$. Thus we learn that there is no enhancement of the isometry group to the conformal group in the Poincaré patch, so we expect that massive theories in the curved space to exhibit some properties of conformal theories. We will explicitly see this later when we compute the mutual information of a massive theory in AdS$_2$ in terms of a conformal cross-ratio, for instance.

Let us introduce a free Dirac fermion $\chi$ in this background. Using the metric \eqref{eq:AdS2metric} the vielbein and spin connection are given by
\be 
e^0=\frac{\mathrm{d}t}{z}\,,\quad e^1=\frac{\mathrm{d}z}{z}\,\quad\text{and}\quad \omega_\mu^{01}=-\frac{\delta_{\mu t}}{z}\,.
\ee
Then the symmetrized action becomes
\be\label{eq:action_Dirac_AdS} 
\mc S = L\int\dd t\int_0^\infty\dd z\off{\frac{i}{2z}\of{\bar\chi\gamma^0\partial_t\chi+\bar\chi\gamma^1\partial_z\chi-\partial_t\bar\chi\gamma^0\chi-\partial_z\bar\chi\gamma^1\chi}-\frac{mL}{z^2}\bar\chi\chi}.
\ee
Of course, we must supplement this action with some suitable boundary condition in order to have a well-posed variational problem. Let us impose
\be\label{eq:BC_AdS} 
\gamma^0\chi|_{z=0}=-\chi|_{z=0}\,,
\ee
which in the context of AdS/CFT literature is known as the standard quantization of the Dirac field, and yields a unitary theory for all (positive) values of the mass \cite{Henningson:1998cd, Giombi:2021cnr,Laia:2011wf}. We will pick the representation $\gamma^0=-\sigma^2$ and $\gamma^1=i\sigma^3$ for the Dirac matrices. Also, we will perform the field redefinition $\psi(t,z)=(L/z)^{1/2}\,e^{i\pi\sigma^1/4}\chi(t,z)$ which renders the action \eqref{eq:action_Dirac_AdS} as
\be \label{eq:accionfermionads}
\mc S=i\int\dd t\int_0^\infty\dd z\ \bar\psi\off{\frac{1}{2}\of{\sigma^3\overset{\leftrightarrow}{\partial}_t+i\sigma^2\overset{\leftrightarrow}{\partial}_z}-\frac{\sigma^1mL}{z}}\psi\, ,
\ee
and the boundary condition to
\be\label{eq:BC_AdS_v2} 
    \sigma^{3} \psi|_{z=0} = -\psi|_{z=0} \, .
\ee
The equation of motion for this fields is
\be\label{eq:EOM} 
\left(\sigma^3\partial_t+i\sigma^2\partial_z-\frac{\sigma^1mL}{z}\right)\psi=0\,.
\ee
This gives a set of two coupled differential equations, one for each spinor component, that can be decoupled by acting on \eqref{eq:EOM} with the conjugate operator. Given the boundary behavior \eqref{eq:BC_AdS_v2}, the solutions to these equations are given by a superposition of the following functions which remain regular at the boundary\footnote{The general solution to the equation \eqref{eq:EOM} includes an extra term which is divergent at $z=0$. However, this term is ruled out by the boundary condition \eqref{eq:BC_AdS_v2} which sets the upper component of the spinor $\psi(t,z)$ to zero at the boundary.}
\be 
\label{eq:AdS_modes}
\psi(t,z)=e^{-ikt}\sqrt{z}\,\binom{J_{mL+1/2}(kz)}{-iJ_{mL-1/2}(kz)}\, .
\ee
We remark that if we identify $z\leftrightarrow r$ and
\be\label{eq:mL_mu}
mL=\mu_\ell\, ,
\ee
the action \eqref{eq:accionfermionads} becomes equivalent to that of the dimensionally reduced theories \eqref{eq:S_lms}. Moreover, the boundary behavior for both models is the same, thus \eqref{eq:AdS_modes} precisely matches the regular radial solutions \eqref{eq:solutions_radial}. This implies that under this map of parameters, the vacuum state and correlation functions of the Dirac fermion in AdS$_2$ will coincide with the corresponding ones for the two dimensional theories appearing in the dimensional reduction of the fermionic model in $d$ dimensions.
Thus, each of these theories are actually equivalent to a Dirac fermion propagating in AdS$_2$ with mass given by the eigenvalue $\mu_\ell$ appearing in \eqref{eq:Z_lms}, in units of the inverse AdS radius.

\subsection{Exact result for the $n$th Rényi mutual information}

We have seen that the theories appearing in the dimensional reduction of the parent fermionic field are equivalent to the standard quantization of Dirac fermions propagating in AdS$_2$ by taking $mL=\mu_\ell$. Particularly, this implies that we can compute the RMIs on the half line \eqref{eq:RMI_half-line} as RMIs on AdS$_2$. In this section, we will show how these quantities can be exactly computed in terms of Painlevé trascendants. Recently, this technology was also used in order to obtain an example for an entropic $C$-function in AdS \cite{Abate:2026apg}.

A standard method for computing the $n$th RMI consists on replicating the theory $n$ times. In this context the $n$th Rényi entropy of a subregion $V$ coincides with the expectation value of a twist operator which applies a permutation of the replicated fields over the region \cite{Cardy:2007mb, Casini:2009sr}. Arranging all the copies of our fields $\psi_{n}$ in a single vector $\vec{\psi}$, the Rényi-twist operator $\Sigma_{n}^{V}$ acts on this field as
\begin{equation}
    \Sigma_{n}^{V}\, \vec{\psi}(x) \of{\Sigma_{n}^{V}}^{\dagger} = \begin{cases}
        T \vec{\psi}(x) &\text{if}\ x\in V \\
        \vec{\psi}(x) &\text{if}\ x\notin V
    \end{cases}\,,
\end{equation}
where $T$ is the matrix that applies a cyclic permutation. For a free Dirac fermion, it is
\begin{equation}
    T = \begin{pmatrix}
        0 & 1 & 0 & \cdots & 0 \\
        0 & 0 & 1 & \cdots & 0 \\
        \vdots & \vdots & \vdots & \ddots & 1 \\
        (-1)^{n+1} & 0 & 0 & \cdots & 0
        \end{pmatrix} \, ,
\end{equation}
where the factor $(-1)^{n+1}$ picked up between the first and last copy comes from the fermionic statistics.

We can diagonalize the matrix $T$ with the change of basis $U$. Its eigenvalues are $e^{2\pi ik/n}$ for $k \in \{-(n-1)/2, -(n-1)/2 + 1, \cdots , (n-1)/2 \}$. Let us work with the fields that diagonalize the action of the Rényi-twist operator, $\vec{\xi}(x) = U \vec{\psi}(x)$. Then, the Rényi-twist operators act on every component of the transformed field as a $U(1)$ twist operator $u^V(2\pi k/n)$, namely they apply the $U(1)$ symmetry locally in $V$ for every $\xi_j$ as
\begin{equation}
    u^{V}(2\pi q)\, \xi_{j}(x) \of{u^{V}(2\pi q)}^{\dagger} = \begin{cases}
        e^{2\pi iq} \xi_{j}(x) \ &\text{if} \ x\in V \\
        \xi_{j}(x) \ &\text{if} \ x\notin V
    \end{cases} \, .
\end{equation}
Since the Lagrangian of the theory \eqref{eq:accionfermionads} is invariant under the change of basis $U$, each $\xi_{j}(x)$ is an independent theory, therefore the Rényi-twist operator can be written as
\begin{equation}
    \Sigma_{n}^{V} = \bigotimes_{k = - \frac{n-1}{2}}^{{\frac{n-1}{2}}} u^{V} \left( \frac{2 \pi k}{n} \right) \, .
\end{equation}
Then, the Rényi entropies appearing in \eqref{eq:RMI_half-line} in terms of this twist operators are given by
\begin{equation}
    \expval{\Omega^{\otimes n}}{\Sigma_{n}^{V}}{\Omega^{\otimes n}} = \Tr \left( \rho_{V}^{n} \right) = \mathrm{exp}\,\big((1-n)\, S_{n}(V;\mu_\ell)\big)\ ,
\end{equation}
where $\ket{\Omega^{\otimes n}} = \ket{\Omega} \otimes \overset{n}{\cdots} \otimes \ket{\Omega}$ is the vacuum of the replicated theory and $\rho_{V} = \Tr_{V^{\prime}} \left( \ketbra{\Omega}{\Omega} \right)$ is the reduced density matrix over the region $V$, computed by tracing out the degrees of freedom in the complementary region $V^{\prime}$.
The following combination of expectation values then gives the desired $n$th RMI appearing in \eqref{eq:RMI_half-line},
\begin{equation}
\begin{split}
        \mathrm{exp}\,\big(-(1-n)\, I_{n}(A,B;\mu_\ell) \big) 
        &= \frac{\expval{\Omega^{\otimes n}}{\Sigma_{n}^{A}\Sigma_{n}^{B}}{\Omega^{\otimes n}}}{\expval{\Omega^{\otimes n}}{\Sigma_{n}^{A}}{\Omega^{\otimes n}} \expval{\Omega^{\otimes n}}{\Sigma_{n}^{B}}{\Omega^{\otimes n}}} \\
        &= \prod_{k=-\frac{n-1}{2}}^{\frac{n-1}{2}} \frac{\expval{\Omega}{u^{A} \left( \frac{2 \pi k}{n} \right) u^{B} \left( \frac{2 \pi k}{n} \right)}{\Omega}}{\expval{\Omega}{u^{A} \left( \frac{2 \pi k}{n} \right)}{\Omega}\expval{\Omega}{u^{B} \left( \frac{2 \pi k}{n} \right)}{\Omega}} \ .
\label{eq:RMI_from_twist}
\end{split}
\end{equation}

The expectation value of the $U(1)$ twist operators $u^{V}(2\pi q)$ and their two-point function for a free fermion in an AdS$_2$ background with vanishing boundary conditions have been computed by Palmer et al. and Doyon in \cite{Palmer:1993jp, Doyon:2003nb}, respectively.\footnote{The two point function of twist operators is also known for the flat space case, see \cite{Casini:2005rm}.} These authors studied certain spinless operators $\mc O_q(x)$ with scaling dimension $q^2$ and some specific properties. Namely, whenever the Dirac field goes around them counterclockwise (in Euclidean signature) it acquires a phase given by $e^{2\pi iq}$, so these operators and the Dirac field are not mutually local. This operators are precisely the $U(1)$ twist operators
\begin{equation}
    \mathcal{O}_{q}(a) = u^{A}(2 \pi q) \, ,
\label{eq:twist_scalingfields}
\end{equation}
corresponding to the region $A = (0, a)$. The $U(1)$ twist operator on $B=(b,\infty)$ can also be written as
\begin{equation}
    u^{B}(2 \pi q) = u^{B^{\prime}}(-2 \pi q)U(2 \pi q) \ ,
\end{equation}
where $U(2 \pi q)$ is the global operator applying the $U(1)$ symmetry, $B^{\prime} = (0,b)$ and now we can identify $u^{B^{\prime}}(-2 \pi q)$ with $\mathcal{O}_{-q}(b)$ through \eqref{eq:twist_scalingfields}.
Moreover, since the global vacuum state is $U(1)$ invariant
\begin{equation}
    \expval{\Omega}{u^{A} \left( 2 \pi q \right) u^{B} \left(2 \pi q \right)}{\Omega} 
    = \langle\mc O_{q}(a)\mc O_{-q}(b)\rangle\,.
\end{equation}
In \cite{Palmer:1993jp, Doyon:2003nb} it was shown that this correlator is given by the tau function of a certain Painlevé differential equation, namely
\be\label{eq:DvsEta} 
\langle\Omega| u^A(2\pi q)u^B(2\pi q)|\Omega\rangle=\tau_q(s)\,,\qquad s=\tanh^2\of{\frac{D}{2L}}=1-\eta\,.
\ee
Here $D=\log(b/a)$ is the geodesic length between the endpoints of the intervals $A$ and $B$ located in the bulk of AdS$_2$, and $\eta$ is the same invariant cross-ratio appearing in \eqref{eq:crossratio}. As we anticipated earlier, although we are not dealing with a CFT, the cross-ratio again plays the role of the relevant dimensionless variable. More precisely, the tau function is computed via the following relation
\begin{equation}
\begin{split}
    \frac{\dd}{\dd s} \log(\tau_q(s))
    &=  \frac{s(1-s)}{4w(1-w)(w-s)}\left(w' - \frac{1-w}{1-s}\right)^2
    - \frac{\mu_{\ell}^2}{w-s} \\
    &\qquad+ \frac{q^2 w}{s(1-s)}
    - \frac{q^2}{s}
    + \frac{\mu_\ell^2  - q^2}{(1-s)} \ ,
\label{eq:dlntau_def}
\end{split}
\end{equation}
where $w(s)$ is a solution to the Painlevé VI differential equation
\begin{equation}
\begin{split}
    &w'' 
    - \frac{1}{2}\left(\frac{1}{w} + \frac{1}{w-1} + \frac{1}{w-s}\right)(w')^2
    + \left(\frac{1}{s} + \frac{1}{s-1} + \frac{1}{w-s}\right) w' \\
    &\qquad\qquad\qquad= \frac{w(w-1)(w-s)}{s^2(1-s)^2} \left(
        \frac{\of{1-4\mu_\ell^2}s(s-1)}{2(w-s)^2}
        -\frac{s}{2w^2}
        + 2 q^{2}
    \right) ,
\label{eq:DE_painleve}
\end{split}
\end{equation}
matching the following boundary conditions in the limit $s\to1$
\begin{equation}\label{eq:BC_painleve}
    w(s) = 1 + \alpha_q\of{\mu_\ell^2-q^2}(1-s)^{1 + 2 \mu_\ell} s^{2 q} \ _{2}F_{1}(1 + q + \mu_\ell, q + \mu_\ell, 1 + 2\mu_\ell,1-s)^{2} \, ,
\end{equation}
with
\begin{equation}\label{eq:alpha_k}
    \alpha_q=\of{ 
    \frac{
    \sin(\pi q) \,
    \Gamma(1+\mu_\ell-q)\,
    \Gamma(1+\mu_\ell+q)
    }{
    \pi\Gamma(2+2\mu_\ell)
    }}^2.
\end{equation}
See \cite{Doyon:2003nb} for more details.

When the separation between $A$ and $B$ is large, clustering decomposition implies that the correlator must factorize. Since this limit is attained by taking $\eta\to0$, we expect
\begin{equation}
    \tau_{q}(1) = \expval{\Omega}{u^{A} \left( 2 \pi q \right)}{\Omega} \expval{\Omega}{u^{B} \left( 2 \pi q \right)}{\Omega} \, .
\end{equation}
Therefore, we can express the ratio between expectation values as
\begin{equation}\label{eq:correlator_ratio}
    \frac{\expval{\Omega}{u^{A} \left( 2 \pi q \right) u^{B} \left( 2 \pi q \right)}{\Omega}}{\expval{\Omega}{u^{A} \left( 2 \pi q \right)}{\Omega}\expval{\Omega}{u^{B} \left( 2 \pi q \right)}{\Omega}} = \exp{ - \int_{1-\eta}^{1}  \frac{\dd}{\dd s} \log\of{\tau_{q}(s)}\dd s} \, .
\end{equation}
So defining the function
\begin{equation}
    v_{q}(\eta)= \int_{1-\eta}^{1} \frac{\dd}{\dd s} \log\of{\tau_{q}(s)} \dd s \, ,
\label{eq:vdef}
\end{equation}
which is explicitly calculable by means of \eqref{eq:dlntau_def}, we arrive to a formula for the $n$th Rényi mutual information between the intervals $A$ and $B$ for a free Dirac field in AdS$_2$ by comparing \eqref{eq:correlator_ratio} to \eqref{eq:RMI_from_twist}. Explicitly, we have 
\begin{equation}
    I_{n}(A,B;\mu_\ell) =\frac{1}{1-n} \sum_{k=-\frac{n-1}{2}}^{\frac{n-1}{2}} v_{k/n}(\eta) \, .
\label{eq:RenyiMutual_ads}
\end{equation}

\begin{figure}[h]
    \centering
    \includegraphics[width=0.9\linewidth]{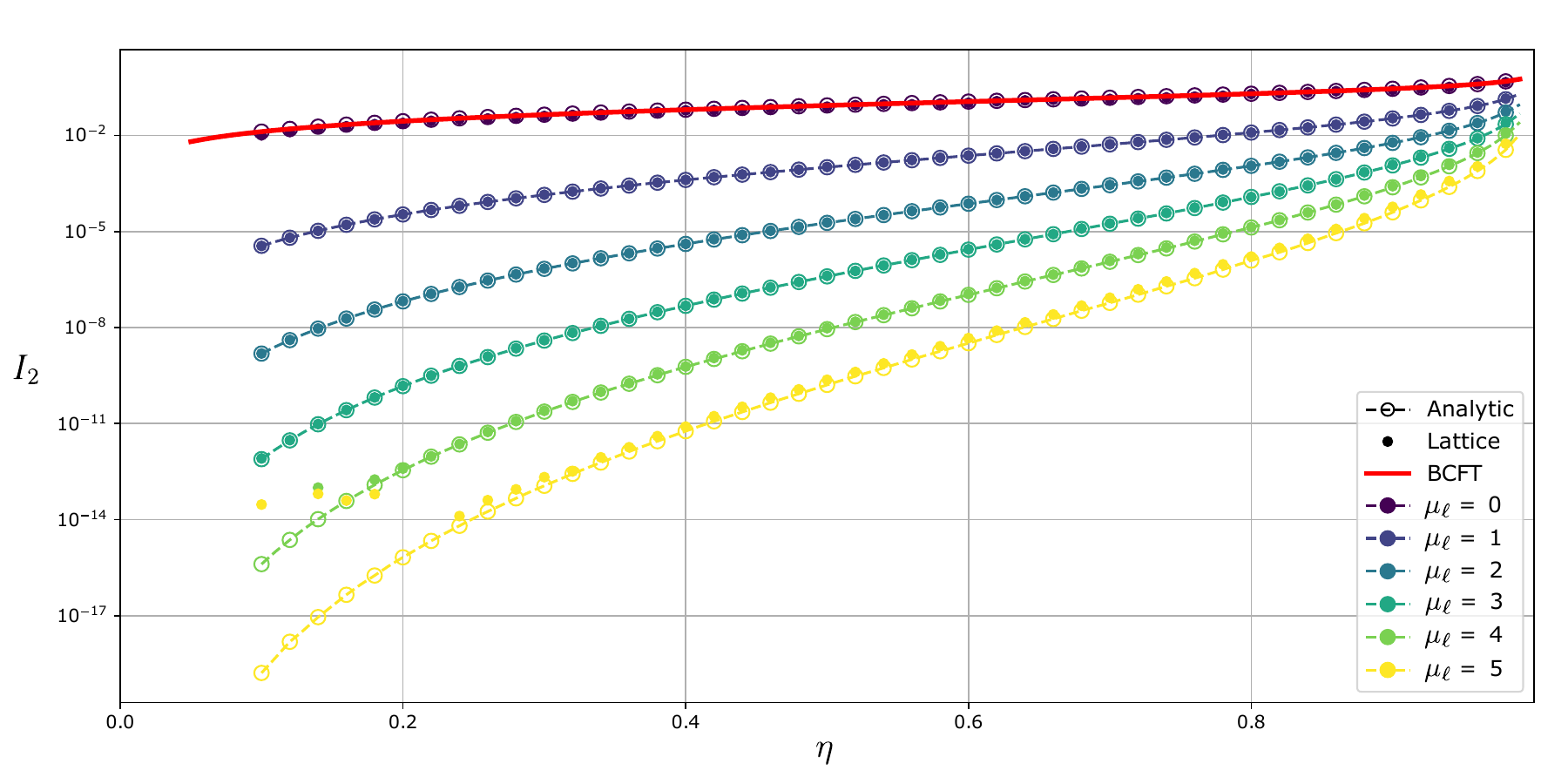}
    \caption{Numerical comparison of the Rényi mutual information for a Dirac field in AdS$_2$, given by \eqref{eq:RenyiMutual_ads} (empty dashed circles), with a lattice calculation using a discretized version of the Hamiltonian of the dimensional reduced theory \eqref{eq:hamiltonian_dimReg_fermion} (filled circles) for several values of the parameter $\mu_{\ell}$ and $n = 2$.
    The red line represents the exact result for $\mu_{\ell}=0$, which is the case of a boundary conformal field theory, obtained in \cite{Mintchev:2020uom}.
    }
    \label{fig:LatticevsFermion_n2}
\end{figure}

In Fig. \ref{fig:LatticevsFermion_n2}, we plot \eqref{eq:RenyiMutual_ads} for $n=2$ and several values of $\mu_{\ell}$. Similar results are obtained for other values of $n$. When $\eta\to0$ there is an exponential suppression with increasing values of the parameter $\mu_\ell$. On the other hand, for $\eta\to1$ all the curves lie within the same order. We will delve into this later when we study the asymptotical behavior of the RMIs and the sum over angular momentum modes. We also compare the results with a lattice calculation using a discretized version of the Hamiltonian of the dimensional reduced theory \eqref{eq:hamiltonian_dimReg_fermion} (See Appendix \ref{app:LatticeComputations} for details of the implementation). We find good agreement between the two approaches, with small deviations coming from numerical limitations in the implementation of the lattice computation or in the solving of the Painleve VI equation. Moreover, we also show the analytic result for the conformal case $\mu_\ell=0$, which was obtained in \cite{Mintchev:2020uom} using BCFT techniques. 

\subsection{Analytic continuation and mutual information}\label{sec:analytic_cont}

In this section, we perform the analytic continuation in the parameter $n$ of the equation \eqref{eq:RenyiMutual_ads}, taking inspiration from the analytic continuation used for the Rényi mutual information of the chiral current studied in \cite{Abate:2025ywp, Arias:2018tmw}. Ultimately, we are interested in obtaining an expression for the limit $n \rightarrow 1$.

\begin{figure}[h]
    \centering
    \includegraphics[width=0.5\linewidth]{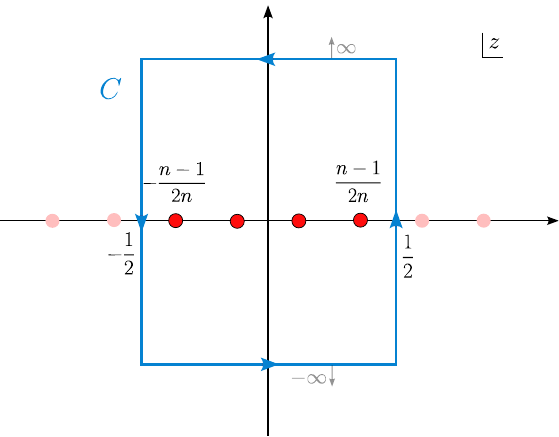}
    \caption{Representation of the contour of integration $C$, which encircles the poles of \eqref{eq:tan} that lie within $-1/2\leq\mathrm{Re}(z)\leq1/2$ (opaque red dots in the Figure). The paths parallel to the real axis are taken to $\pm \infty$ respectively.}
    \label{fig:contour_int}
\end{figure}

For the analytic continuation we propose the ansatz
\begin{equation}
    I_{n}(\eta)
    =  \frac{i n }{2(1-n)} \oint_{C} \dd z \ v_{z}(\eta) \left[ \tan\big(n \pi (z+1/2)\big) - \tan\big(\pi (z+1/2)\big) \right] ,
\label{eq:analytic1}
\end{equation}
for the rectangular contour $C$ shown in Fig. $\ref{fig:contour_int}$. When $n\in\mathbb{N}$, using the residue formula one can show that this integral reproduces the equation \eqref{eq:RenyiMutual_ads} due to the structure of singularities of the function 
\begin{equation}\label{eq:tan}
    - n \pi \tan\big(n \pi (z+1/2)\big) = \sum_{l\in\mathbb{Z}} \of{z - \frac{1}{n}\of{l - \frac{n-1}{2}}}^{-1} \,,
\end{equation}
whose poles located within the interior of $C$ are precisely the values for $k/n$ that occur in the sum \eqref{eq:RenyiMutual_ads}. Substracting the term $\tan(\pi (z+1/2))$ adds a pole at $z = 0$, but its contribution is suppressed because $v_{0}$ vanishes for $\eta \in [0,1]$. Note that, in this case, the $U(1)$ twist operators simply become the identity operator.\footnote{From the point of view of the differential equation, the normalization constant \eqref{eq:alpha_k} for $q = 0$ gives $\alpha_q = 0$, then the solution to \eqref{eq:DE_painleve} is $w(s) = 1$ and thus $\frac{\dd}{\dd s} \log(\tau_{0}(s)) = 0$.}

As one pushes the horizontal segments in $C$ towards infinity the term $\tan(\pi(z+1/2))$ becomes fundamental, since it ensures an exponential damping for the integrand in the imaginary directions. Namely,
\begin{equation}
    \lim_{t \rightarrow \pm \infty} \left[\tan\big(n \pi (s  + it)\big) - \tan\big(\pi (s + it)\big) \right] \propto e^{\mp 2 \pi c t} \rightarrow 0 \qquad \text{for} \ s \in \mathbb{R} \ \text{and} \ n  > 0 \, ,
\end{equation}
where $c = \text{min}\,(1, n)$. Here, we are also assuming that $v_{z}$ does not grow exponentially fast in the imaginary direction. Then, the only non-vanishing contributions to the integral just come from  the vertical segments crossing the real line at $\pm 1/2$. Performing the integral over those paths we get  
\begin{equation}
\begin{split}
    z = 1/2 + it\quad \rightarrow \quad
    & i \int_{-\infty}^{\infty} \dd t \left[ \tan\big(n \pi (1+it)\big) - \tan\big(\pi (1+it)\big) \right]v_{\frac{1}{2} + it}(\eta) \\
    & \qquad\qquad\quad= - \int_{-\infty}^{\infty} \dd t\ \big(\! \tanh(n \pi t) - \tanh(\pi t) \big)\,v_{\frac{1}{2} + it}(\eta)\,, \\ 
    z = - 1/2 - it \quad\rightarrow \quad
    & -i \int_{-\infty}^{\infty} \dd t \left[ \tan\big(n \pi (-it)\big) - \tan\big(\pi (-it)\big) \right]v_{- \frac{1}{2} - it}(\eta) \\
    & \qquad\qquad\quad= - \int_{-\infty}^{\infty} \dd t\ \big(\! \tanh(n \pi t) - \tanh(\pi t) \big)\,v_{- \frac{1}{2} - it}(\eta)\,.
\end{split}
\end{equation}
Using that $v_{-z}(\eta) = v_{z}(\eta)$, which can be seen from \eqref{eq:dlntau_def} and \eqref{eq:DE_painleve}, we arrive at an analytic continuation of the $n$th RMI expressed as an integral over the real axis, which resembles that of the chiral scalar current in $1+1$ dimensions \cite{Arias:2018tmw}
\begin{equation}
    I_{n}(A,B;\mu_\ell)
    = \frac{i n }{n-1} \int_{-\infty}^{\infty} \dd t \ \big(\!\tanh(n \pi t) - \tanh(\pi t) \big)\, v_{\frac{1}{2} + i t}(\eta) \,.
\label{eq:RMIext}
\end{equation}

From this analytic continuation we can safely recover the mutual information by taking the limit $n \rightarrow 1$. Noting that
\begin{equation}
    \lim_{n \rightarrow 1}\ \frac{1}{n-1} \big(\!  \tanh(n \pi t) - \tanh(\pi t) \big) = \frac{\pi t}{\cosh^{2}(\pi t)} \,,
\end{equation}
and replacing in \eqref{eq:RMIext} we get the following
\begin{equation}\label{eq:MI_integral}
    I(A,B;\mu_\ell) = \lim_{n \rightarrow 1} I_{n}(A,B;\mu_\ell)
    = i \int_{-\infty}^{\infty} \dd t \ \frac{\pi t}{\cosh^{2} \left( \pi t \right)}\,v_{\frac{1}{2} + i t}(\eta) \, .
\end{equation}
Actually, this resembles the expression given in \cite{Casini:2009sr} for the entanglement entropy and $C$-function of a massive Dirac fermion in flat space. In order to explicitly see this, we rewrite our integral as
\begin{equation}
    I(A,B;\mu_\ell)
    = -i \int_{\Upsilon} \dd z \ \frac{\pi (z - \frac{1}{2})}{\sin^{2} \left( \pi z \right)} \,v_{z}(\eta) \, ,
\label{eq:integralMIv2}
\end{equation}
where $\Upsilon$ is the contour parametrized by $z = 1/2 + i t$ with $t \in \mathbb{R}$, which is the right boundary of the strip $0 < \Re(z) \leq 1/2$. The integrand is holomorphic inside the strip and goes to zero exponentially for $\Im(z) \rightarrow \pm \infty$, therefore we can safely move the contour to the left boundary, located at $z = it$. Thus we get
\begin{equation}
    I(A,B;\mu_\ell) 
    = \int_{-\infty}^{\infty} \dd t \ \frac{\pi (it - \frac{1}{2})}{\sin^{2} \left( \pi it \right)}\, v_{i t}(\eta) 
    = \int_{0}^{\infty} \dd t \ \frac{\pi}{\sinh^{2} \left( \pi t \right)}\,v_{-i t}(\eta) 
    \ ,
\label{eq:integralMI}
\end{equation}
where we used again that $v_{-z}(\eta) = v_{z}(\eta)$. This is the final expression we will use for computing mutual informations in our work. A useful rule for the following is the identification
\begin{equation}\label{eq:cont_nto1}
    \frac{1}{1-n} \sum_{k = - \frac{n-1}{2}}^{\frac{n-1}{2}} f(k/n)
    \ \overset{n  \rightarrow 1}{\longrightarrow} 
    \ \int_{0}^{\infty} \dd\kappa \ \frac{\pi }{\sinh^{2} \left( \pi \kappa \right)} f(- i \kappa) \, ,
\end{equation}
which can be inferred by comparing \eqref{eq:RenyiMutual_ads} and our final expression \eqref{eq:integralMI}.

\begin{figure}[h]
    \centering
    \includegraphics[width=0.95\linewidth]{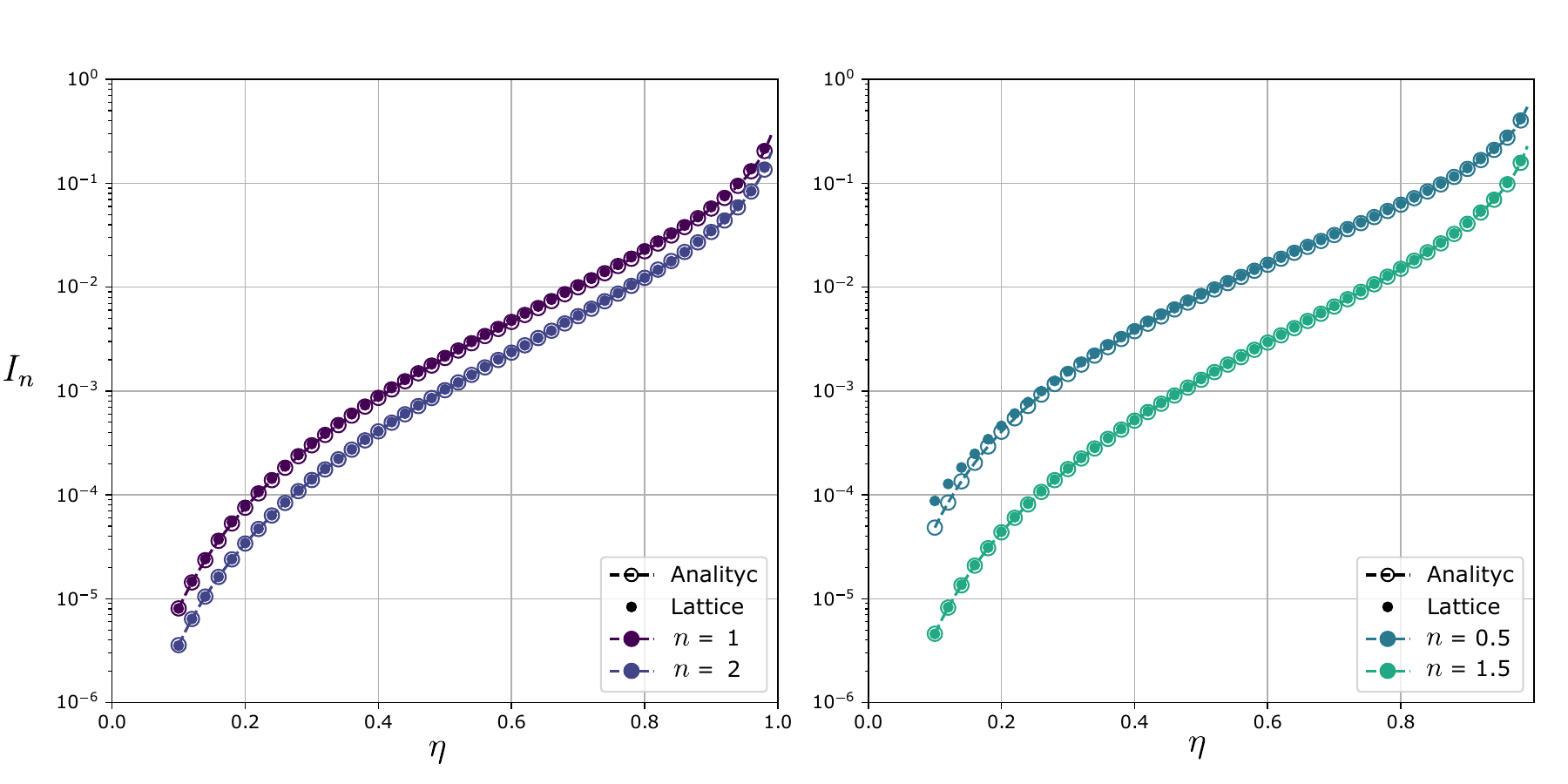}
    \caption{Numerical comparison of the Rényi mutual information for a Dirac field in AdS$_2$ (empty dashed circles), given by \eqref{eq:RenyiMutual_ads} for $n = 2$, by \eqref{eq:integralMI} for $n = 1$ and by its analytic continuation \eqref{eq:RMIext} for non-integer $n$, with a lattice calculation using a discretized version of the Hamiltonian of the dimensional reduced theory \eqref{eq:hamiltonian_dimReg_fermion} (filled circles) for several values of the parameter $n$. In all cases we have set $\mu_\ell=1$.}
    \label{fig:LatticevsFermion_m1}
\end{figure}

Although we assumed some properties for $v_z$ as a function of $z$ in the intermediate steps of our derivation without providing a rigorous proof, we checked that our formula is correct by comparing with simulations on the lattice. In Fig. \ref{fig:LatticevsFermion_m1} we plot the RMI for several values of $n$, where we used \eqref{eq:RenyiMutual_ads} for $n=2$, \eqref{eq:integralMI} for $n=1$ and \eqref{eq:RMIext} for $n\not\in\mathbb{N}$, and we compare it with the results for the lattice (see Appendix \ref{app:LatticeComputations} for details of the lattice implementation). There, we can see a good agreement between both approaches, validating our analytical results.

\subsection{Asymptotics}

In this section we will consider the limits when the intervals $A$ and $B$ are very far apart or very close to each other. The former implies taking $a\ll b$, and thus $\eta\to0$, meanwhile the latter is $a\simeq b$, or equivalently $\eta\to1$. We will use these results in the next section when comparing to, and also expanding the knowledge about the expected behavior for the $d$ dimensional parent theory, but we will also see that this analysis on the half line is already interesting on its own.

\subsubsection{Long distance expansion}

Let us first address the limit when $\eta\to0$. An asymptotical expression for $v_{k/n}$ was given in \cite{Doyon:2003nb} for this regime, which actually translates in the boundary condition \eqref{eq:BC_painleve} used to find solutions to the Painlevé equation. In terms of the AdS geodesic distance between the endpoints of the intervals, this is
\be\label{eq:doyon_asymptotics} 
v_{k/n}(D)\simeq-\log\of{1+4^{2\mu_\ell+1}\alpha_{k/n}e^{-(2\mu_\ell+1)D/L}+4^{2\mu_\ell+2}\beta_{k/n}e^{-(2\mu_\ell+2)D/L}},
\ee
plus higher order terms,\footnote{The higher order corrections are explicitly computed in Appendix B in \cite{Doyon:2003nb}.} where $\alpha_{q}$ was given in \eqref{eq:alpha_k} and
\be
\beta_{q}=\frac{q^2}{2}\frac{2\mu_\ell+1}{(\mu_\ell+1)^2}\,\alpha_{q}\,.
\ee
Expansion \eqref{eq:doyon_asymptotics} holds whenever $D\gg L$ regardless of the value of the mass. Note that the exponential decay persists even at $\mu_\ell=0$, which is a typical feature of field theories in AdS where the radius acts as an additional scale. Thus, we can infer that the condition
\be\label{eq:corr_length} 
 D \gg \frac{L}{2\mu_\ell+1}\sim\frac{L}{\mu_\ell}\,
\ee
is controlling the long distance asymptotics, acting as a correlation length. Moreover, this is consistent with the picture on the half line, where the position-dependent mass $\mu_\ell/r$ appearing in the action for each angular momentum mode \eqref{eq:S_lms} implies that the correlation length of the theory must be given by some characteristic scale divided by the parameter $\mu_\ell$. 

We can find $D$ as a function of $\eta$ inverting the relation \eqref{eq:DvsEta}. This gives
\begin{equation}\label{eq:geodesic_dist}
    D = L \,\log \left( \frac{1 + \sqrt{1 - \eta}}{1 - \sqrt{1 - \eta}} \right),
\end{equation}
so we can write \eqref{eq:doyon_asymptotics} as a expansion in powers of $\eta$, carefully keeping track of the contributions coming from both exponential functions. First, one expands the argument inside the logarithm on \eqref{eq:doyon_asymptotics} up to the power $\eta^{2\mu_\ell+2}$ yielding
\begin{equation}
\begin{aligned}
&1+4^{2\mu_\ell+1}\alpha_{k/n}e^{-(2\mu_\ell+1)D/L}+4^{2\mu_\ell+2}\beta_{k/n}e^{-(2\mu_\ell+2)D/L}
\\
&\qquad\qquad\qquad=1+\alpha_{k/n}\eta^{2\mu_\ell+1}+\big(\beta_{k/n}+ \tfrac{1}{2} (2 \mu_\ell+1)\,\alpha_{k/n}\big)\eta^{2\mu_\ell+2}+\mc O(\eta^{2\mu_\ell+3}) \, .
\end{aligned}
\end{equation}
Then we use the following formula for the expansion of the logarithm
\be
\log\of{1+c_px^p+c_{p+1}x^{p+1}}\simeq\begin{cases}
    c_px^p+c_{p+1}x^{p+1} & \text{for}\ p>1\\
    c_1x+(c_2-c_1^2/2)x^2 & \text{for}\ p=1
\end{cases}\,,
\ee
that replacing in \eqref{eq:RenyiMutual_ads} gives
\be
\label{eq:In_largeDist_perL}
I_n(A,B;\mu_\ell)=-\frac{1}{1-n}\off{\of{\sum_k\alpha_{k/n}}\eta^{2\mu_\ell+1}+\of{\sum_k\tilde\beta_{k/n}}\eta^{2\mu_\ell+2}}+\mc O\of{\eta^{2\mu_\ell+3}}\,,
\ee
where
\be
\tilde\beta_{k/n}=\begin{cases}\beta_{k/n}+ \tfrac{1}{2}(2\mu_\ell+1)\,\alpha_{k/n}& \text{for}\quad \mu_\ell\neq0\\
\beta_{k/n}+\tfrac{1}{2}(1-\alpha_{k/n})\,\alpha_{k/n}& \text{for}\quad \mu_\ell=0
\end{cases}\ .
\ee
Here, we can explicitly see the exponential suppression of the RMI due to the parameter $\mu_{\ell}$ in the limit $\eta \to 0$, as we remarked when introducing Fig. \ref{fig:LatticevsFermion_n2}.

The $n\to1$ limit of \eqref{eq:In_largeDist_perL} can be easily obtained by the replacement \eqref{eq:cont_nto1}. This gives
\begin{align}\label{eq:I_largeDist_perL} 
I(A,B;\mu_\ell)&=\frac{\sqrt{\pi}\,\Gamma(2\mu_\ell+2)}{2^{4\mu_\ell+3}\Gamma(2\mu_\ell+5/2)}\,\eta^{2\mu_\ell+1}+\dbtilde{\beta}(\mu_\ell)\,\eta^{2\mu_\ell+2}+\mc O\of{\eta^{2\mu_\ell+3}},
\end{align}
where
\be
\dbtilde{\beta}(\mu_\ell)=-\int_0^\infty\dd\kappa\ \frac{\pi \tilde\beta_{-i\kappa}}{\sinh^2(\pi\kappa)}\,,
\ee
which involves computing integrals of powers of gamma functions that can be easily integrated numerically.

It is worth noting that we obtained a formula which resembles the long distance behavior of RMIs for CFTs, even though we are dealing with a massive theory. Actually, the exponent of the leading term in \eqref{eq:In_largeDist_perL} is precisely twice the scaling dimension of the Dirac field near the boundary of AdS, as can be seen from \eqref{eq:AdS_modes}, where the theory acquires conformal invariance. We find this connection very interesting and would like to further explore it in the future.

\subsubsection{Short distance expansion}

In the opposite limit, the two-point function of the Rényi-twist operators acquires a conformal behavior
\begin{equation}\label{eq:conformal}
    \expval{\Omega}{ u^{A}(2 \pi q) u^{B}(2 \pi q)}{\Omega} \simeq D^{-2q^2},
\end{equation}
in terms of the geodesic distance between the endpoints of the intervals $D$. This is the UV limit of the model. It is worth stressing that in order for this approximation to hold \eqref{eq:corr_length} suggests that, in contraposition, we need
\be\label{eq:short_condition} 
D \ll \frac{L}{2 \mu_\ell + 1} \,.
\ee
A first step to achieve this is to take the intervals $A$ and $B$ close to each other, namely $b=a+\epsilon$ with $\epsilon\ll a$. In this setup, one has $D/L=\log(1+\epsilon/a)\simeq\epsilon/a$, so condition \eqref{eq:short_condition} translates into a stronger bound for the separation between the intervals when $\mu_{\ell}\gg1$,
\be\label{eq:short_condition_epsilon} 
\epsilon\ll a/\mu_\ell\,.
\ee
Let us analyze this from the perspective on the half line. Recall we are studying correlations between the regions $A$ and $B$, and these are strongest near their boundaries $a\simeq b$. There, the mass term in the action \eqref{eq:S_lms} takes the value $\mu_\ell/a$, and then condition \eqref{eq:short_condition_epsilon} on $\epsilon$ just comes naturally from comparing it with the inverse of the mass.

In the following, let us assume \eqref{eq:short_condition_epsilon} holds. Then, one gets $\eta\to1$ and $D \simeq 2L\, \sqrt{1 - \eta}$ in this limit. Therefore, we have the following formula for the ratio of expectation values of twist operators,
\begin{equation}\label{eq:1ptfunc}
    \frac{\expval{\Omega}{u^{A}(2 \pi q) u^{B}(2 \pi q)}{\Omega}}{\expval{\Omega}{u^{A}(2 \pi q)}{\Omega} \expval{\Omega}{u^{B}(2 \pi q)}{\Omega}} \simeq \frac{1}{\expval{\Omega}{u^{A}(2 \pi q)}{\Omega}^{2}} \frac{1}{(2L)^{2 q^{2}} (1 - \eta)^{q^{2}}} \, .
\end{equation}
Note that in order to proceed we need an expression for the one-point function of the Rényi-twist operators. This was explicitly calculated in \cite{Doyon:2003nb} as the following integral
\begin{equation}
    \expval{\Omega}{u^{A}(2 \pi q)}{\Omega} = (2L)^{-q^{2}} \exp{\int_{0}^{\infty} \frac{\dd t}{t} \of{1 - e^{-2 \mu_\ell t}} \frac{\sinh^{2}(q t)}{\sinh^{2}(t)} } ,
\end{equation}
so replacing in \eqref{eq:1ptfunc} and further in \eqref{eq:RMI_from_twist}, one gets the short distance limit ($\eta \rightarrow 1$) for the RMI for the theories with mass parameter $\mu_\ell$ on the half line
\begin{equation}
    I_{n}(A,B;\mu_\ell) \simeq - \frac{1+n}{12 n} \log \left( 1 - \eta \right) + \frac{2}{1-n} \sum_{k = -\frac{(n-1)}{2}}^{\frac{(n-1)}{2}} \int_{0}^{\infty} \frac{\dd t}{t} \of{1 - e^{-2 \mu_\ell t}} \frac{\sinh^{2}(kt/n)}{\sinh^{2}(t)}\, .
\label{eq:RenyiMutualShortDistance}
\end{equation}
If we replace $\mu_\ell = 0$ the second term vanishes and we recover the result obtained in \cite{Mintchev:2020uom} for the massless Dirac field on the half line. Curiously, we arrived at this exact result using a short distance approximation.

\begin{figure}[h]
    \centering
    \includegraphics[width=1\linewidth]{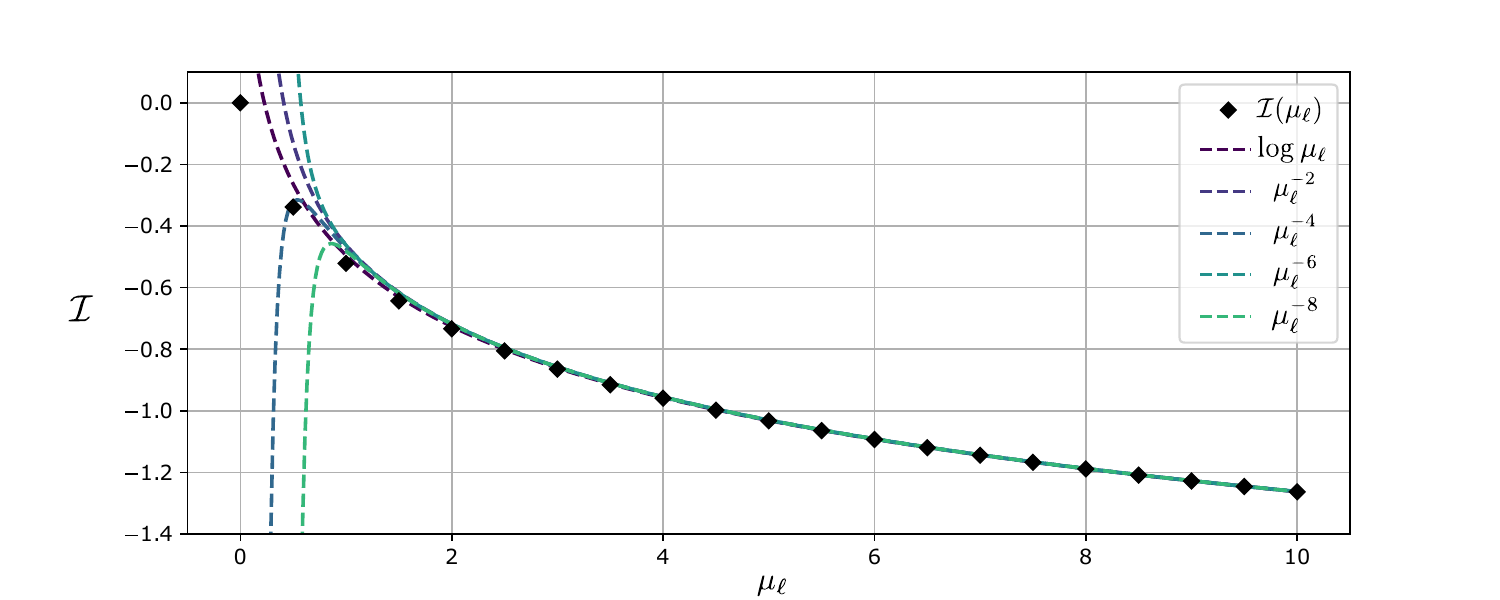}
    \caption{Comparison between some exact values of $\mc I(\mu_\ell)$ and its large $\mu_\ell$ expansion \eqref{eq:expansion_I(mu)}. We see that the approximation is excellent starting from relatively small values of $\mu_\ell$.}
    \label{fig:I(mu)}
\end{figure}

In order to take the limit $n \rightarrow 1$, we will perform the shortcut \eqref{eq:cont_nto1} to replace $k/n \rightarrow - i\kappa$ and the sum with the integral
\begin{equation}
    \int_{0}^{\infty} \dd\kappa\  \frac{\pi \sinh^{2}(-i\kappa t)}{\sinh^{2}(\pi \kappa)} = \frac{1 - t \coth(t)}{2} \, .
\end{equation}
We then get the short distance limit of the mutual information
\begin{equation}\label{eq:mutual_shortdist}
    I(A,B;\mu_\ell) \simeq - \frac{1}{6} \log \left( 1 - \eta \right) + \int_{0}^{\infty} \frac{\dd t}{t} \of{1 - e^{-2 \mu_\ell t}}\frac{1-t \coth(t)}{\sinh^{2}(t)} \,.
\end{equation}
Note that all the geometrical dependence appears in the first term, while the second term in this expression,
\begin{equation}
\begin{aligned}
    \mc I(\mu_\ell) &= \int_{0}^{\infty} \frac{\dd t}{t}\of{1 - e^{-2 \mu t}} \frac{1-t \coth(t)}{\sinh^{2}(t)} \\
        &= \frac{1}{2} \int_{-\infty}^{\infty} \dd t\ \frac{1 - e^{-2 \mu_\ell |t|}}{t}\, \text{sign}(t) \, \frac{1 - t \coth(t)}{\sinh^{2}(t)} \, ,
\end{aligned}
\end{equation}
collects all the dependence in the parameter $\mu_\ell$. Although we were not able to solve the integral analytically, we can approximate it for sufficiently large $\mu_\ell$, always keeping in mind condition \eqref{eq:short_condition_epsilon} on $\epsilon$ holds. In order to simplify its expression, we will consider the integrand as a product of two functions and then work with their Fourier transforms. Namely, let
\begin{equation}
\begin{split}
    G(t) &= \frac{1 - e^{-2 \mu_\ell |t|}}{t}\, \text{sign}(t)\ \rightarrow\ \hat{G}(k) = \log \left( 1+\frac{4 \mu_\ell^{2}}{k^{2}} \right) , \\
    F(t) &= \frac{1 - t \coth(t)}{\sinh^{2}(t)}\ \rightarrow\ \hat{F}(k) = - \frac{\pi^{2} k^{2} }{4 \sinh^2(\pi k/2)} \, .
\end{split}
\end{equation}
By Parseval identity we have that
\begin{equation}
    \mc I(\mu_\ell) = - \frac{1}{4 \pi} \int_{-\infty}^{\infty} \dd k \ \log \left( 1+\frac{4 \mu_\ell^{2}}{k^{2}} \right) \frac{\pi^{2} k^{2} }{4 \sinh^2(\pi k/2)} \, .
\end{equation}
From here, we can easily extract a logarithmic divergence in $\mu_\ell$ as follows
\begin{align}\label{eq:I(mu)}
    \mc I(\mu_\ell) &= - \frac{1}{4 \pi} \int_{-\infty}^{\infty} \dd k \ \log \left( 1+\frac{4 \mu_\ell^{2}}{k^{2}} \right) \frac{\pi^{2} k^{2} }{4 \sinh^2(\pi k/2)} \ , \notag\\
    &= - \frac{1}{4 \pi} \int_{-\infty}^{\infty} \dd k \left[ 2 \log \left( \mu_\ell \right) - \log \left( \frac{k^{2}}{4} \right) + \log \left( \frac{k^{2}}{4 \mu_\ell^{2}} + 1 \right) \right] \frac{\pi^{2} k^{2} }{4 \sinh^2(\pi k/2)} \ , \notag\\
    &= - \frac{1}{3} \log  \mu_\ell  + c_{0} - \frac{1}{4 \pi} \int_{-\infty}^{\infty} \dd k\  \log \left( \frac{k^{2}}{4 \mu_\ell^{2}} + 1 \right) \frac{\pi^{2} k^{2} }{4 \sinh^2(\pi k/2)}   \ ,
\end{align}
where $c_{0} \approx - 0.495$. Clearly, the remaining integral is bounded for $\mu_\ell \rightarrow \infty$. We  expand it in powers of $\mu_\ell^{-2}$ using that
\begin{equation}
    \log \left( 1 + \frac{k^{2}}{4 \mu_\ell^{2}} \right) = - \sum_{j=1}^{\infty} \frac{(-1)^{j}}{j} \left(\frac{k^{2}}{4} \right)^{j} \frac{1}{\mu_\ell^{2 j}} \ ,
\label{eq:expan_log}
\end{equation}
and
\begin{equation}
    \int_{-\infty}^{\infty}\dd k\ \frac{\pi^{2} k^{2j} }{4 \sinh^2(\pi k/2)} %= \frac{2 (2j)! \zeta(2j)}{\pi^{2j-1}}
    = (-1)^{j-1} 2^{2j} \pi B_{2j}\ ,
\end{equation}
where $B_{j}$ is the $j$th Bernoulli number. Replacing in \eqref{eq:I(mu)}, we arrive at
\begin{equation}\label{eq:expansion_I(mu)}
    \mc I(\mu_\ell) \simeq -\frac{1}{3} \log \mu_\ell  + c_{0} + \sum_{j=1}^{\infty} \frac{B_{2j+2}}{j} \frac{1}{\mu_\ell^{2j}}\,.
\end{equation}
This series turns out to be asymptotic because we replace the expansion \eqref{eq:expan_log}, which requires that $k^{2}/4 \mu_{\ell} \ll 1$, in the integral \eqref{eq:I(mu)} where $k^{2}$ takes arbitrary large values. However, this asymptotic series approximates the function in the limit $\mu \rightarrow \infty$, and we check that the first few terms already provide an excellent approximation for $\mu_\ell\geq1$ as we show in Fig. \ref{fig:I(mu)}. Replacing this on \eqref{eq:mutual_shortdist} and explicitly writing the first two contributions in \eqref{eq:expansion_I(mu)} we get

\begin{equation}
    I(A,B;\mu_\ell) \simeq - \frac{1}{6} \log( 1 - \eta) - \frac{1}{3} \log \mu_\ell  + c_{0} - \frac{1}{30\mu_\ell^2} + \frac{1}{84\mu_\ell^4} + \cdots \ .
\label{eq:shortdistancemubig}
\end{equation}
Here we explicitly see what we anticipated in Fig. \ref{fig:LatticevsFermion_n2}: in contraposition with the long distance asymptotics, in the short distance case we have a much slower decay in the mutual information as $\mu_\ell$ increases. This representation of the short distance expansion will be useful for future calculations.

When the intervals are almost complementary, the mutual information famously serves as a geometrical regulator for the entanglement entropy between $A$ and its complement \cite{Casini:2015woa}, namely $I(A,B;\mu_\ell)\simeq2S(A;\mu_\ell)$. So, replacing $b= a+\epsilon$ in \eqref{eq:shortdistancemubig} we have
\be\label{eq:EE} 
2S(A;\mu_\ell)=\frac{1}{3}\log\frac{a}{\epsilon}-\frac{1}{3}\log\mu_\ell-\frac{1}{30\mu_\ell^2}+\frac{1}{84\mu_\ell^4}+\dots,
\ee
where we dropped the constant $c_0$ since it can be absorbed in the definition of $\epsilon$. This entropy has been already computed in \cite{Huerta:2023dqt}, where a similar structure was found. There the authors consider both helicities together, which gives twice the entropy for each one, so we can directly compare to \eqref{eq:EE}. Explicitly, under the identification $\mu_\ell=\ell+(d-2)/2$ and up to an irrelevant constant, they write
\be\label{eq:EE_MG} 
2S(A;\mu_\ell)=\frac{1}{3}\log\frac{a}{\epsilon}+f(\mu_\ell)\,.
\ee
where $f(\mu_\ell)$ is some known integral that cannot be solved analytically. We compared both results for some finite values of $\mu_\ell$ yielding the same results. Moreover, an expansion for $f$ in the large $\mu_\ell$ regime is provided, which turns out to coincide with the one we obtained. So we can confidently claim that our results are equivalent.\footnote{A fast way to check the equivalence of the large $\mu_\ell$ expansion is to replace $\ell \rightarrow \mu_\ell$ and $d \rightarrow 2$ in eqs. (81)-(85) of \cite{Huerta:2023dqt}. Then, one precisely gets \eqref{eq:EE}.}

\section{Mutual information in $d$ dimensions}\label{sec:d_dimensions}

In this section we will combine the results of Sec. \ref{sec:dimReg} and \ref{sec:AdS} to arrive at the RMI and mutual information for the massless Dirac fermion in $d$ dimensions and the Rarita-Schwinger field in $d=4$ across spherical regions. Moreover, we will explore the short and long limits for the separation distance between the spheres using the asymptotical results obtained previously for each value of $\mu_\ell=\ell+(d-2)/2$. 

\subsection{Multipolar expansion}

So far, in \eqref{eq:RMI_sum} we have established that the RMI for a massless Dirac fermion in $d$ spacetime dimensions between two spheres $\mc A$ and $\mc B$ with radius $a$ and $b$, respectively, can be computed as a sum over angular momentum modes,
\be
I_n^{(d)}(\eta)=\sum_{\ell = 0}^{\infty} 2\lambda(\ell)I_n(A,B;\mu_\ell)\,,
\ee
where $I_n(A,B;\mu_\ell)$ is the RMI between the intervals $A=(0,a)$ and $B=(b,\infty)$ for a Dirac fermion on the half line with mass term $\mu_\ell/r$. Moreover, in Sec. \ref{sec:AdS} we showed that the theories on the half line are equivalent to Dirac fermions in AdS$_2$ with mass $\mu_\ell$ in units of the inverse AdS radius, where the RMI can be exactly computed using \eqref{eq:RenyiMutual_ads}. So, putting all togheter,
\be\label{eq:multipolarRMI_final} 
I_n^{(d)}(\eta)=\sum_{\ell = 0}^{\infty} \frac{2\lambda(\ell)}{n-1}\sum_kv_{k/n}(\eta,\mu_\ell)\,,
\ee
where we highlighted the dependence of $v_{k/n}$ with the mass parameter $\mu_\ell$. For the $n\to1$ limit, using \eqref{eq:integralMI} gives the mutual information
\be\label{eq:multipolarMI_final}
I^{(d)}(\eta) = \sum_{\ell = 0}^{\infty} 2\lambda(\ell)\int_0^\infty\dd\kappa\ \frac{\pi v_{-i\kappa}(\eta,\mu_\ell)}{\sinh^2(\pi\kappa)}\,.
\ee

\begin{figure}
    \centering
    \subfigure[]{\includegraphics[width=0.48\linewidth]{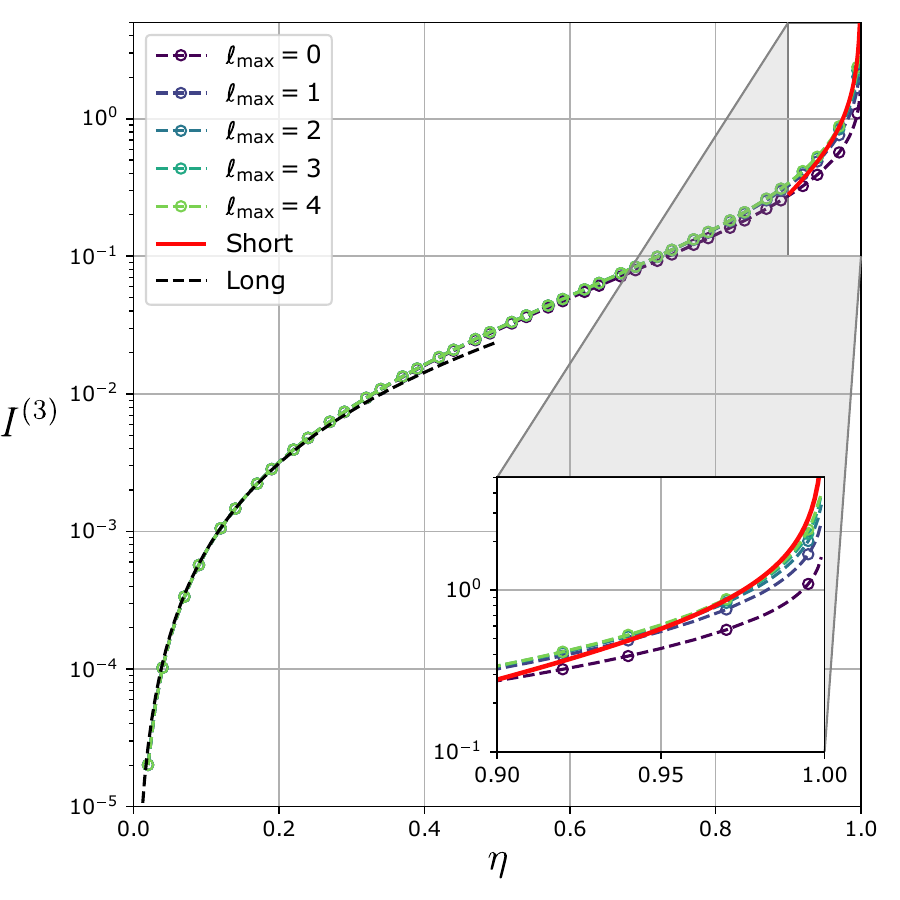}}
    \subfigure[]{\includegraphics[width=0.48\linewidth]{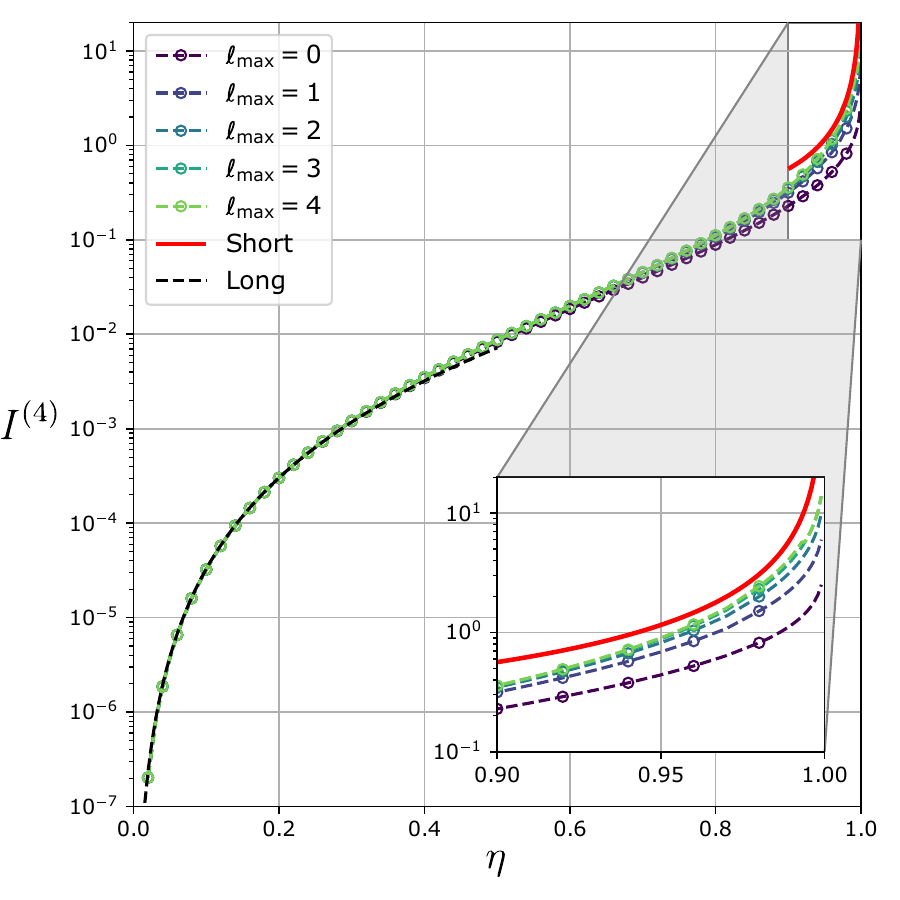}}
    \caption{
        Plot of the partial sums of \eqref{eq:multipolarMI_final} for different values of $\ell_{\text{max}}$, which give and approximation for the mutual information in (a) $d=3$ and (b) $d=4$. The black and red curves represent the ``long'' (as shown below in \eqref{eq:I_4Dirac}-\eqref{eq:I_3Dirac}) and ``short''  distance approximations \eqref{eq:I_3d_short}-\eqref{eq:I_4d_short}, respectively.
    }
    \label{fig:partial_sums}
\end{figure}

One can calculate partial sums of \eqref{eq:multipolarMI_final}, up to some finite $\ell_{\text{max}}$, to get the mutual information in dimensions $d=3$ and $d=4$, as we show in Fig. \ref{fig:partial_sums}. Note that the sum converges rapidly for small $\eta$, which is expected since in the long distance approximation each mode is exponentially suppressed with $\mu_\ell$ as discussed around \eqref{eq:In_largeDist_perL}. The result of using this approximation for $\eta\to0$ is also shown in the plot, and will be explained in detail below. 

In the limit $\eta\to1$, we expect that the mutual information approaches (twice) the entanglement entropy for the sphere $\mc A$, that in $d=3$ and $d=4$ is given by 
\begin{align}
    \label{eq:I_3d_short}
    I^{(3)}\simeq2S^{(3)}(a)
    &= 2s_3\, \frac{a}{\epsilon} + 2F_3
    = \frac{s_3}{\sqrt{1 - \eta}} + 2F_3\,,\\
    \label{eq:I_4d_short}
    I^{(4)}\simeq2S^{(4)}(a)
    &= 2s_4 \left( \frac{a}{\epsilon} \right)^{2} - \frac{11}{45} \log  \frac{a}{\epsilon} 
    = \frac{s_4}{2 (1 - \eta)} + \frac{11}{90} \log \big( 4 (1 - \eta)  \big)\,.
\end{align}
Here $\epsilon=b-a\ll a$ and $s_d=2^{[(d-2)/2]}\mathrm{Area}(\mathbb{S}_{d-2})\kappa_d$, where  $\kappa_d$ is the universal area contribution to the mutual information between parallel plates \cite{Casini:2009sr,Casini:2015dsg} as we will soon review, and also $F_3=-0.2189$ \cite{Klebanov:2011gs}. In Fig. \ref{fig:partial_sums} we see that, although the convergence is much slower, the partial sums do approach \eqref{eq:I_3d_short} and \eqref{eq:I_4d_short} by increasing $\ell_{\max}$ in the limit $\eta \to 1$.

There is an interesting case were we can simply omit the sum over $\ell$. Recall that if the sum begins at $\ell=1$ instead, \eqref{eq:multipolarRMI_final} gives the RMI for the Rarita-Schwinger field in $d=4$ dimensions. Thus, we can explictily compute the difference between the RMI for the Rarita-Schwinger and the Dirac field in $d=4$: it is simply given by the $\ell=0$ (i.e. $\mu_\ell=1$) contribution. Calling this difference $\Delta I_n$, we have then
\begin{align} 
\Delta I_n(\eta)&=\frac{4}{n-1}\sum_k v_{k/n}(\eta,1)\,,\\
\Delta I(\eta)&=4\int_0^\infty\dd\kappa\ \frac{\pi v_{-i\kappa}(\eta,1)}{\sinh^2(\pi\kappa)}\,.
\end{align}
This expressions, albeit given in terms of transcendental functions, can be evaluated numerically yielding an exact result for every $\eta\in(0,1)$. In the left pannel of Fig. \ref{fig:LatticevsFermion_m1}, we have already shown these quantities for $n=1$ and $n=2$. Moreover, the short distance expansion \eqref{eq:mutual_shortdist} gives the difference of the entanglement entropies between these theories for spheres
\be 
\Delta S(a)=\frac{1}{6}\log\frac{a}{\epsilon}+\mc I(1)\,.
\ee
This explicitly coincides, up to an irrelevant additive constant, with the expression proposed in \cite{Benedetti:2023jye} for this entropy difference.

\subsection{Long distance and CFT data}

Let us address the behavior of the RMI when the spheres $\mc A$ and $\mc B$ are very far apart. In this configuration, we expect the mutual information for the $d$ dimensional theory to exhibit some typical CFT behavior. Namely, we should obtain a power law in terms of the cross ratio $\eta$, where the exponents are twice the dimension of the primary operators of the theory \cite{Calabrese:2010he,Cardy:2013nua}. We recall that from \eqref{eq:corr_length} and the fact that $\mu_{\ell} \ge 0$, we could estimate that the long-distance expansion works in $d$ dimensions if
\begin{equation}
    \log \left( 1 + \frac{l}{a} \right) \gg 1 \,,
\label{eq:condition_long}
\end{equation}
holds, where $l=b-a$ is the radial distance bewteen the spheres.

Expressions \eqref{eq:In_largeDist_perL} and \eqref{eq:I_largeDist_perL} are useful for obtaining the long distance expansion of the parent $d$ dimensional theory. Crucially, since each term comes with a power of $2\ell$, we can obtain the first two leading contributions for the RMI just by studing the lowest $\ell$ sector. Let us do so in the following. 

\paragraph{Dirac field:} For the spin 1/2 theory in any dimension, the first terms then just come from the $\ell=0$ contribution.\footnote{Note that the first two orders for $d=4$ then will coincide with those from the expansion for $\Delta I_n$, but of course both quantities differ at higher orders since $I_n^{(4)}$ also gets contributions from the $\ell\geq1$ sectors.} The relevant parameter is then $\mu_0=(d-2)/2$, and replacing in \eqref{eq:multipolarRMI_final} and \eqref{eq:multipolarMI_final} yields
\begin{align}\label{eq:RMI_longDist} 
I_n^{(d)}(\eta)&=\frac{2^{[d/2]}}{n-1}\off{\of{\sum_k\alpha_{k/n}}\eta^{d-1}+\of{\sum_k\tilde\beta_{k/n}}\eta^{d}} + \mc O\of{\eta^{d+1}},\\
\label{eq:MI_longDist}
I^{(d)}(\eta)&= 2^{[d/2]}\off{\frac{\sqrt{\pi}\,\Gamma(d)}{2^{2d-1}\Gamma(d+1/2)}\,\eta^{d-1}+\dbtilde{\beta}\big((d-2)/2\big)\,\eta^{d}}  + \mc O \of{\eta^{d+1}} ,
\end{align}
plus terms of order $\eta^{d+1}$ and higher which mix with the $\ell>0$ sectors.

When $d$ is even, the coefficients of the RMI can be obtained as a rational function of $n$ using \eqref{eq:RMI_longDist}, which can later be safely continued to $n\to1$. Meanwhile, for odd $d$ we have not obtained a closed form for the sum over $k$, but the mutual information can still be computed using \eqref{eq:MI_longDist}. Since in any case the general expressions are rather lengthy, let us focus on the $n=1$ case here and leave the $n\neq1$ case for App. \ref{app:longDist}. At $d=3$ and $d=4$ we have
\begin{align}\label{eq:I_4Dirac} 
I^{(4)}(\eta)&=\frac{1}{35}\eta^3+\frac{4}{105}\eta^4+\mc O\of{\eta^5},\\
\label{eq:I_3Dirac} 
I^{(3)}(\eta)&=\frac{1}{15}\eta^2+\frac{2}{35}\eta^3+\mc O\of{\eta^4}.
\end{align}
In Fig. \ref{fig:partial_sums} we show how \eqref{eq:I_4Dirac} and \eqref{eq:I_3Dirac} describe the mutual information for $\eta \rightarrow 0$ very accurately, in the region where the partial sums already converged. 

Regarding $d=2$, the $\ell=0$ sector actually gives the full contribution and there is no need for the factor of 2 accounting for both the $\pm$ modes. This time the explicit $n$ dependence is quite simple,
\be\label{eq:In_2Dirac} 
I_n^{(2)}(\eta)=\frac{1 + n}{12 n}\eta+\frac{1+n}{24n}\eta^2\,+\mc O\of{\eta^3},
\ee
and we recover the Taylor expansion for the expression given in \cite{Mintchev:2020uom}, as expected. The leading terms in \eqref{eq:I_4Dirac}-\eqref{eq:In_2Dirac} precisely match those computed in \cite{Chen:2017hbk}. More generically, the leading term in \eqref{eq:MI_longDist} coincides with the expression given in \cite{Casini:2021raa} for arbitrary dimensions. Moreover, our expressions also yield the leading contributions for arbitrary $n$, and further give access to subleading terms. To the best of our knowledge, these results are not present in the literature up to the date.

Let us also remark that we obtained the expected conformal behavior in any spacetime dimension. For a Dirac field the lowest scaling dimension operator is the field itself, with scaling dimension $\Delta=(d-1)/2$. Thus the exponent of the leading term should be $2\Delta=d-1$, and this is precisely what happens in our expansion. 

\paragraph{Rarita-Schwinger field:} We can also use \eqref{eq:RMI_longDist} or \eqref{eq:MI_longDist} for the spin 3/2 case in $d=4$, where this time the first two terms in the expansion will come from the $\ell=1$ sector. Again let us leave the expression for the RMI to App. \ref{app:longDist} and show the result for $n=1$ here,
\be\label{eq:I_RS} 
I^{(4)}(\eta)=\frac{2}{693}\eta^5+\frac{20}{3003}\eta^6+\mc O\of{\eta^7}.
\ee
This is another novel result of our work. In this case, the lowest scaling dimension operator that we should consider is not the field itself, but the field strength $F_{\mu\nu}=\partial_\mu\Psi_\nu-\partial_\nu\Psi_\mu$ instead, which is gauge invariant and whose dimension is $\Delta=5/2$. Thus, we can identify the correct power for the leading term in our result.

\subsection{Short distance and universal terms}

In the remainder of this section, we will show how \eqref{eq:multipolarMI_final} gives access to some specific terms in the mutual information for the parent theory in the $\eta\to1$ limit, which are naturally inherited by the entanglement entropy.  For instance in $d=4$, this analysis will reproduce both contributions appearing in \eqref{eq:I_4d_short}.

\subsubsection{Area}

\begin{figure}[h]
    \centering
    \includegraphics[width=0.5\linewidth]{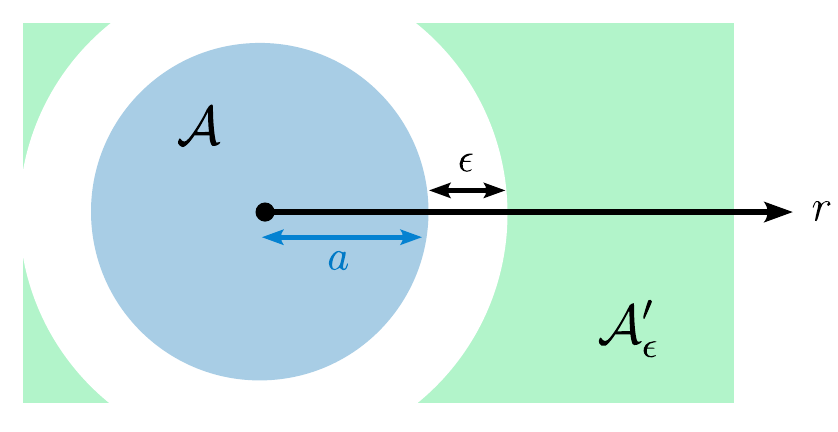}
    \caption{Geometrical setup used to extract the area term in the mutual information, where $a\gg\epsilon$.}
    \label{fig:area}
\end{figure}

For any dimension $d$, the leading term in the $\eta\to1$ limit of the mutual information is known to be proportional to the area of the sphere once inherited by the entanglement entropy. More concretely, taking $\epsilon=b-a$ small, one has
\be\label{eq:area_term} 
I^{(d)}\big(\mc A,\mc A'_\epsilon)=2^{[d/2]}\mathrm{Area}(\mathbb{S}_{d-2})\,\kappa_d \of{\frac{a}{\epsilon}}^{d-2}+\mc O\of{(a/\epsilon)^{d-4}},
\ee
where we introduced the notation $\mc A'_\epsilon$ for the region which is almost complementary to $\mc A$, as shown in Fig. \ref{fig:area}. The coefficient $\kappa_d$ is exactly known to be
\be \label{eq:kappa}
\kappa_d=\of{(d-2)\,2^{d-3}\pi^{(d-2)/2}\,\Gamma\of{\frac{d-2}{2}}}^{-1}\int_0^\infty\dd t\ t^{d-3}c_{\mathrm{flat}}(t)\,,
\ee
with $c_{\mathrm{flat}}(t)$ the entropic $C$-function for a single massive fermionic degree of freedom in $d=2$ Minkowski space. See Table 2 in \cite{Casini:2009sr} for a list with explicit values of this coefficient in several dimensions.

Let us show how this same result arises using our multipolar expansion. First, we will approximate the sum over the angular momentum as an integral, because the divergence will come from the terms with large $\ell$. Also, since $\lambda(\ell)$ is a polynomial in $\ell$, we can just keep the term with the highest order to extract the leading divergence we are after. This gives   
\be
I^{(d)}(\mc A,\mc A'_\epsilon)=\frac{2^{[d/2]}}{(d-3)!}\int_0^\infty\dd\ell\ \ell^{d-3}I(A,A'_\epsilon;\mu_\ell)\,,
\label{eq:mutual_d_integral}
\ee
plus subleading corrections, where $A=(0,a)$ and $A'_\epsilon=(a+\epsilon,\infty)$. 

We want to analyze the limit where $\epsilon \rightarrow 0$ which by conformal invariance is the same as $a \rightarrow \infty$. Coming back to the picture in AdS, we have that the geodesic distance in this limit is given by
\begin{equation}
    D = L \log\of{\frac{a+\epsilon}{a}} \simeq L\, \frac{\epsilon}{a} \ . 
\end{equation}
If we take $L \propto a \rightarrow \infty$ we are simultaneously taking the flat space limit, where the geodesic distance goes to the flat space distance $D \rightarrow \epsilon$.
In this limit, under the transformation $r\to r-a$, we have that the mutual information $I(A,A'_\epsilon;\mu_\ell)$ in AdS goes to the mutual information between the semi-infinite intervals $(-\infty, 0)$ and $(\epsilon, \infty)$ for a Dirac fermion of mass $m = \lim_{a \rightarrow \infty}\mu_{\ell}/a$ in flat space, recalling \eqref{eq:mL_mu}. Namely,
\begin{equation}
    I(A,A^{\prime}_ \epsilon; \mu_{\ell})       \overset{a \rightarrow \infty}{\longrightarrow} I_{\text{flat}}(\epsilon,m) \,.
\label{eq:mutual_limit_flat}
\end{equation}
We note that by translation invariance in flat space, we have
\begin{equation}
    \epsilon \partial_{\epsilon} I_{\text{flat}}(\epsilon,m) = - \epsilon \partial_{\epsilon} S_{\text{flat}}(\epsilon, m) = -\widetilde{c}_{\text{flat}}(m\epsilon)\,,
\end{equation}
where $S_{\text{flat}}(\epsilon,m)$ is the entanglement entropy of an interval of length $\epsilon$ for a massive Dirac field, and $\widetilde{c}_{\text{flat}}(t)$ is the $C$-function associated to it, which only depends on $\epsilon$ trough the dimensionless combination $t = m \epsilon$ \cite{Casini:2005rm}. Therefore, the mutual information $I_{\text{flat}}(\epsilon, m)$ also only depends on $\epsilon$ trough $t$.
Thus, going back to \eqref{eq:mutual_d_integral} and taking the limit $a \rightarrow \infty$ as we explained before, considering $\ell \sim \mu_{\ell} = m a$ and using \eqref{eq:mutual_limit_flat} we arrive at
\be\label{eq:I_area_I} 
I^{(d)}(\mc A,\mc A'_\epsilon)=\frac{2^{[d/2]}}{(d-3)!}\of{\frac{a}{\epsilon}}^{d-2}\int_0^\infty\dd t\ t^{d-3}I_{\mathrm{flat}}(t)\, .
\ee
Moreover, integrating by parts 
\be
\begin{split}
\int_0^\infty\dd t\ t^{d-3}I_{\mathrm{flat}}(t)
    &=\of{d-2}^{-1}\of{t^{d-2}I_{\mathrm{flat}}(t)\big|_{t=0}^{t\to\infty}-\int_0^\infty\dd t\ t^{d-2}\partial_t I_{\mathrm{flat}}(t)}\, \\
    &= (d - 2)^{-1} \int_{0}^{\infty} \dd t \  t^{d-3}\, \widetilde{c}_{\text{flat}}(t) \ .
\end{split}
\ee
The boundary terms vanish because of the ``short'' and ``long'' distance behavior of the $\widetilde{c}_{\text{flat}}(t)$ function studied in \cite{Casini:2005rm}, which translate into $I_{\text{flat}}(t) \propto \log(t)$ for $t \approx 0$ and $I_{\text{flat}}(t) \propto t^{-1/2} e^{-2t}$ for $t \rightarrow \infty$.
This $C$-function, which is computed for a Dirac fermion, is twice the one appearing in \eqref{eq:kappa}, which is that of a Majorana fermion. Finally, we arrive at
\be\label{eq:I_area_c} 
I^{(d)}(\mc A,\mc A'_\epsilon)=\frac{2^{[d/2]+1}}{(d-2) (d-3)!}\of{\frac{a}{\epsilon}}^{d-2}\int_0^\infty\dd t\ t^{d-3} c_{\mathrm{flat}}(t)\, .
\ee
Using that $\mathrm{Area}(\mathbb{S}_{d-2})=2\pi^{\frac{d-1}{2}}/\Gamma\big(\frac{d-1}{2} \big)$ and that $(d-3)!=2^{d-3}\pi^{-1/2}\,\Gamma\big(\frac{d-2}{2}\big)\Gamma\big(\frac{d-1}{2}\big)$, one precisely gets the leading term in \eqref{eq:area_term}, recovering the area contribution to the $d$ dimensional parent theory.

\subsubsection{Logarithmic contribution}

When $d$ is even, the mutual information also acquires a logarithmic divergence with $\epsilon$ on top of the area contribution and other subleading terms. The logarithmic coefficient coincides with (twice) the type-A trace anomaly of the Dirac fermion in $d$ dimensions. Given its dependence with $\epsilon$, once inherited by the entanglement entropy, this coefficient becomes cutoff independent unlike the other contributions. In \cite{Huerta:2023dqt}, the logarithmic coefficient for the $d$ dimensional Dirac fermion was obtained from the dimensionally reduced theories using \eqref{eq:EE_MG} and introducing an ad hoc exponential regulator. In the following we will rederive their result but using the mutual information instead, treating the correlation length in the half line as the natural cutoff.

Recall that \eqref{eq:short_condition_epsilon} controls the short and long distance asymptotics of the mutual information. That is, given some fixed $\epsilon=b-a$, there is a characteristic value for the angular momentum $\ell_*$ such that
\be 
\mu_{\ell_*}\sim\frac{a}{\epsilon}\,,
\ee
and such that when $\ell\ll\ell_*$ the mutual information on the half line is well approximated by the short distance expansion \eqref{eq:mutual_shortdist}, while for $\ell\gg\ell_*$, it is well approximated by the long distance expansion \eqref{eq:I_largeDist_perL}. Ultimately, we will take the limit $\epsilon\to0$ or equivalently $\ell_*\gg1$. With  this in mind, we will split the sum over $\ell$ as
\be 
I^{(d)}(\eta)=\sum_{\ell=0}^{\ell_*}2\lambda(\ell)I_{\mathrm{short}}(A,B;\mu_\ell)+\sum_{\ell>\ell_*}2\lambda(\ell)I_{\mathrm{long}}(A,B;\mu_\ell)\,,
\ee
where the subscripts ``short'' and ``long'' refer to the approximation we will use for each summand. Since in the long distance case the mutual information is exponentially suppressed with $\ell$, while the degeneracy $\lambda(\ell)$ is polynomial, we expect that these contributions will converge fast enough, yielding a finite result. Whereas, in the short distance case, the sum will converge much slower, strongly depending on the value of $\epsilon$. Since we are interested in extracting a divergent $\log\epsilon$ contribution, we will simply drop the long distance sum, so we will focus on the terms with $\ell<\ell_*$. 

Writing
\be 
2\lambda(\ell)=\sum_{q=0}^{d-3}\lambda_q\ell^q\qquad\text{for}\qquad \lambda_q=\frac{2^{d/2}}{(d-3)!}\begin{bmatrix}
    d-2\\q+1
\end{bmatrix}\,,
\ee
for even $d$, where $\big[\begin{smallmatrix} s \\ r \end{smallmatrix}\big]$ are the unsigned Stirling numbers of the first kind, we have
\be 
I^{(d)}\simeq\lambda_0\of{\frac{1}{3}\log\frac{a}{\epsilon}+\mc I\big((d-2)/2\big)}+\sum_{\ell=1}^{\ell_*}\sum_{q=0}^{d-3}\lambda_q\ell^q\of{\frac{1}{3}\log\frac
{a}{\epsilon}+\mc I\big(\ell+(d-2)/2\big)}\,.
\ee
Note that we have isolated the $\ell=0$ contribution, which only involves the constant term in $\lambda(\ell)$. Moreover, we can approximate $\mc I(\mu_\ell)$ by its large $\mu_\ell$ expansion \eqref{eq:expansion_I(mu)} which is already very accurate for $\mu_\ell\sim1$, see Fig. \ref{fig:I(mu)}. This gives
\be\label{eq:I_sums}
I^{(d)}\simeq\frac{\lambda_0}{3}\log\frac{a}{\epsilon}+\sum_{q=0}^{d-3}\frac{\lambda_q}{3}\of{S_1(q)\log\frac{a}{\epsilon}-S_3(q,\tfrac{d-2}{2})}+\sum_{q=0}^{d-3}\sum_{j=1}\frac{\lambda_qB_{2j+2}}{j}S_2(q,2j,\tfrac{d-2}{2})\,,
\ee
where we have dropped constant contributions and introduced the sums
\begin{align}
    S_1(q)&=\sum_{\ell=1}^{\ell_*}\ell^q=\zeta(-q)-\zeta(-q,\ell_*+1)\,,\\
    S_2(q,p,\delta)&=\sum_{\ell=1}^{\ell_*}\frac{\ell^q}{\of{\ell+\delta}^p}=\sum_{j=0}^q\binom{q}{j}(-\delta)^{q-j}\off{\zeta(p-j,\delta)-\zeta(p-j,\ell_*+\delta+1)}\,,\\
    S_3(q,\delta)&=\sum_{\ell=1}^{\ell_*}\ell^q\log\of{\ell+\delta}=\partial_q S_1(q)+\sum_{j=1}\frac{(-1)^{j+1}\delta^j}{j}S_1(q-j)\,,\,
\end{align}
for $\zeta(k,w)=\sum_{j=0}^\infty(j+w)^{-k}$ the Hurwitz Zeta function, satisfying $\zeta(k,0)=\zeta(k)$. We are interested in the limit $w\gg1$ of the Hurwitz Zeta, where this function and its derivative behave differently depending on the value of their first argument, namely
\be 
\zeta(k,w)=\begin{cases}
    \log w+\mc O(w^{-1}) & \text{for}\ k=1\\
    w^{1-k}/(k-1)+w^{-k}/2+\sum_{j\geq1}B_{2j}/(2j)!(k)_{2j-1}w^{1-k-2j} & \text{for}\ k\neq1
\end{cases}\,.
\ee
Moreover, when $k$ is a zero or a negative integer, the sum truncates and $\zeta(k,w)$ becomes a polynomial in $w$. This means that $S_1(q)\log a/\epsilon$ in \eqref{eq:I_sums} will give a logarithm times a polynomial in $\ell_*$, plus a pure logarithmic contribution. Meanwhile $S_2$ will give a logarithm only when $j=p-1$. Regarding the derivative appearing in $S_3$, one has
\be 
\partial_z\zeta(z,w)|_{z=-q}=-\zeta(-q,w)\log w + \mc O\of{w^{q+1}},
\ee
which also gives a logarithm times a polynomial.

Therefore, replacing in \eqref{eq:I_sums} we get the following structure of divergent terms in the limit $\epsilon \rightarrow 0$
\begin{equation}
\begin{split}
    \sum_{\ell=0}^{\ell_*}2\lambda(\ell)I_{\mathrm{short}}
    &= b_{0} \log \of{\frac{a}{\epsilon}} + c_{0} \log \of{\ell_{*}} 
    + \cdots
    + \of{b_{d-3} + c_{d-3} \log \of{ \frac{\ell_{*} \epsilon}{a}}} \ell_{*}^{d-3} \\
    &\qquad+ \of{b_{d-2} + c_{d-2} \log \of{ \frac{\ell_{*} \epsilon}{a}}} \ell_{*}^{d-2}
    \, ,
\end{split}
\label{eq:resum_structure}
\end{equation}
where $b_{j}, c_{j} \in \mathbb{R}$. Furthermore, \eqref{eq:resum_structure} suggests that for obtaining the expected structure for the mutual information in this limit we have to take
\begin{equation}
    \ell_{*} \propto \frac{a}{\epsilon} \, .
\end{equation}
Using this, we get
\begin{equation}
\begin{split}
    I^{(d)}(\eta) &=  c_{\text{log}}^{(d)} \log \of{\frac{a}{\epsilon}}
    + \cdots
    + \# \of{\frac{a}{\epsilon}}^{d-3}
    + \# \of{\frac{a}{\epsilon}}^{d-2}
    \ .
\end{split}
\end{equation}
We remark that all the coefficients in this approximation depend on the proportionality constant between $\ell_{*}$ and $a/\epsilon$, except for the logarithmic one, as expected. Carefully tracking all contributions, we get the following expression for the logarithmic coefficient of the mutual information in arbitrary (even) $d$ dimensions
\be
\begin{split}
c_{\log}^{(d)}&=\frac{2^{d/2}}{(d-3)!}\Bigg\{\!\begin{bmatrix}d-2\\1\end{bmatrix}\frac{2\zeta(0)-d+4}{6}+\sum_{q=1}^{d-3}\begin{bmatrix}d-2\\q+1\end{bmatrix}\Bigg(\frac{\zeta(-q)}{3}+\frac{\big(\!-\!(d-2)/2\big)^{q+1}}{3(q+1)}\\
&\qquad\qquad\qquad\qquad+\sum_{j=1}^{[(q+1)/2]}\binom{q}{2j-1}\frac{B_{2j+2}}{j}\of{-\frac{d-2}{2}}^{q+1-2j}\Bigg)\Bigg\}\,.
\end{split}
\ee
This gives twice the type-A trace anomaly for the Dirac field \cite{Cappelli:2000fe} for even $d$ and vanishes for odd $d$. In Table \ref{tab:c_log} below we explicitly show this for several dimensions.

\begin{table}[h!]
    \centering
    \begin{tabular}{ccccclcccc}\toprule
       $d$  &  $3$& $4$ &$5$ &$6$ & 7&8 & 9&10 &$\dots$\\\midrule
         $c_{\log}^{(d)}$&  0&  $-\frac{11}{45}$& 0&$\frac{191}{1890}$ & 0& $-\frac{2497}{56700}$& 0&$\frac{14797}{748440}$ &$\dots$\\ \bottomrule
    \end{tabular}
    \caption{Logarithmic coefficient for the mutual information for a massless Dirac field in $d$ dimensions, coinciding with twice the given in \cite{Cappelli:2000fe}, as expected.}
    \label{tab:c_log}
\end{table}

\section{Conclusions and outlook}\label{sec:conclusions}

In this article we proposed a novel approach for computing in an exact way the $n$th RMI between spherical regions for free, massless fermionic theories in arbitrary dimensions. Profiting of the spherical symmetry, we dimensionally reduce the $d$ dimensional theories into an infinite tower of two dimensional models on the half line, corresponding to each angular momentum mode. In turn, we show that these two dimensional theories are actually Dirac fermions propagating in AdS$_2$. Using results by Palmer et al. and Doyon \cite{Palmer:1993jp,Doyon:2003nb}, the RMIs with $n\in\mathbb{N}_{>1}$ in AdS$_2$ can be computed in terms of solutions of a certain Painlevé differential equation. Moreover, we obtained an analytical continuation in the Rényi parameter $n$ which ultimately let us compute the $n\to1$ limit of the RMI, namely the mutual information. We also studied the asymptotical limit of the RMIs for the two dimensional theories in the regime where the spheres become very close and very far apart from each other. Then, using these approximations we were able to obtain the leading contributions for the long distance expansion of the RMIs as well as the area and logarithmic universal coefficients appearing in the mutual information of the parent theory in $d$ dimensions.

A natural and very interesting future line of work would be to apply this techniques to bosonic fields. The dimensional reduction can also be performed, giving a tower of scalars in the half line. Preliminary analysis seem to indicate that these theories are also scalars propagating in AdS$_2$, so developing the technology to obtain RMIs in this setting would in principle also let us obtain exact expressions for the bosonic case.

As a byproduct of our computation, we obtained an exact expression for the difference between the mutual information for the Dirac and Rarita-Schwinger fields in $d=4$. In order to do so, we used the fact that the dimensional reduction of the latter model coincides with that of the former, but with the first angular mode removed. This generalizes to higher spin the result obtained in \cite{Abate:2025ywp} for the difference between the RMI of the Maxwell and scalar fields in $d=4$, whose dimensional reduction also differ by the zero mode. Moreover, if we were able to adapt the technology employed in this paper to bosonic fields, we could further extend this to gravitons, whose dimensional reduction coincides with that of the Maxwell field without its first contribution.

Finally, it would also be interesting to further develop the study of quantum information measures in AdS. The isometries of this spacetime reduce to the conformal group on its boundary and operators on the bulk can be written in terms of primary operators of the boundary using the so called boundary operator expansion (BOE) \cite{Paulos:2016fap,Loparco:2026fki,DeCesare:2026bor}. Thus, correlation functions for gaussian theories in AdS can be written in terms of boundary two point functions. This would impact the RMI since it is computed as the correlation function between twist operators on the bulk. We saw this explicitly when studying the long distance limit for the RMI between the intervals $A=(0,a)$ and $B=(b,\infty)$ in AdS$_2$, where the lowest scaling dimension of the boundary theory appeared as the exponent of the leading term. We think that this should also be the case for more general regions and for higher dimensional geometries, where the BOE approach combined with standard conformal techniques could be extremely helpful.

\section*{Acknowledgments}

We are deeply thankful to Horacio Casini and Marina Huerta for enlightening discussions and constant encouragement throughout this work. We also thank Gonzalo Torroba for detailed comments on the draft as well as Guido van der Velde, Valentín Benedetti and Lucas Daguerre for discussions. The work of NA and LM is supported by a CONICET PhD fellowhip.

\appendix

\section{Lattice computations on the half line}
\label{app:LatticeComputations}

In this appendix, we detail the numerical implementation used to compute the entanglement entropy and mutual information for a free Dirac fermion on a half line ($r \ge 0$). The methodology follows the dimensional reduction and lattice discretization techniques established in \cite{Huerta:2011qi,Huerta:2023dqt,Casini:2009sr}.

We consider a two-component Dirac fermion spinor $\psi(r) = (u(r), v(r))^\mathsf{T}$ propagating on the half line. The system is governed by a Hamiltonian \eqref{eq:hamiltonian_dimReg_fermion}
\begin{equation}
    H = \frac{i}{2} \int_0^\infty \dd r \off{
    {\psi}^\dagger(r) \of{-\sigma^1\partial_r+\frac{\mu}{r}\,i\sigma^2} \psi(r) - \mathrm{h.c.} 
    }.
\end{equation}
where $r$ denotes the radial distance to the origin.
To evaluate the correlation functions numerically, we discretize the radial coordinate $r$ into a regular lattice with spacing $\delta \equiv 1$ and $N$ sites
\begin{equation}
    r \to j \in \{1, 2, \dots, N\} \ .
\end{equation}
For the spatial derivative we take
\begin{equation}
    \partial_{r} \psi(r) \rightarrow \psi_{j+1} - \psi_{j} \ .
\end{equation}
The continuous spinor fields $u(r)$ and $v(r)$ map to discrete lattice operators $u_j$ and $v_j$ satisfying standard canonical anti-commutation relations. We organize the internal spin and spatial degrees of freedom into a single unified column vector of operators $\vec{c}$ defined as:
\begin{equation}
    \vec{c} = (u_{1}, v_{1}, u_{2}, v_{2}, \cdots, u_{N}, v_{N}) \ .
\end{equation}
Explicitly, the mapping between the component indices and the vector index is given by $c_{2j-1} = u_j$ and $c_{2j} = v_j$ for the $j$-th lattice site.
Using this basis, Hamiltonian can be cast into a compact quadratic form
\begin{equation}
    H = \sum_{l,m} c_{l}^\dagger \mathcal{H}_{l,m} c_m \ ,
\end{equation}
where $\mathcal{H}$ is a $2N \times 2N$ Hermitian matrix with the following components
\begin{equation}
\begin{split}
    &\mathcal{H}_{2i-1, 2j-1} = \mathcal{H}_{2i, 2j} = 0, \quad \\
    &\mathcal{H}_{2i-1, 2j} = i \left[ \frac{\mu}{j} \delta_{i,j} + \frac{1}{2} \left( \delta_{i,j+1} - \delta_{i,j-1} \right) \right], \quad \\
    &\mathcal{H}_{2i, 2j-1} = i \left[ - \frac{\mu}{j} \delta_{i,j} + \frac{1}{2} \left( \delta_{i,j+1} - \delta_{i,j-1} \right) \right] \ .
\end{split}
\end{equation}

The correlator $C_{lm} = \expval{\Omega}{c_{l}c_{m}^{\dagger}}{\Omega}$ is given by
\begin{equation}
    C = \Theta(-\mathcal{H}) \ ,
\end{equation}
where $\Theta$ is the Heaviside step function acting on the eigenvalues of $\mathcal{H}$. Let $V$ be a spatial subregion of the lattice (for example, an interval $V = [j_{\text{min}}, j_{\text{max}}]$). The reduced density matrix $\rho_V = \text{Tr}_{V^{\prime}}\ketbra{\Omega}{\Omega}$ remains Gaussian, meaning it is entirely determined by the restricted correlation matrix $C_{V}$, obtained by keeping only the rows and columns of $C$ corresponding to the degrees of freedom inside $V$.

The von Neumann entanglement entropy $S(V) = -\text{Tr}(\rho_V \log \rho_V)$ and Rényi entropy $S_{n}(V) = \frac{1}{1-n} \log\text{Tr}(\rho_V^{n})$ can be evaluated directly from the eigenvalues of $C_{V}$ using the standard functional trace formula \cite{Casini:2009sr}
\begin{equation}
    S(V) = -\text{Tr} \left[ C_{V} \log C_{V} + (I - C_{V}) \log (I - C_{V}) \right] \ ,
\end{equation}
\begin{equation}
    S_{n}(V) = \frac{1}{1-n} \text{Tr} \left[ \log \left( C_{V}^{n} + (1-C_{V})^{n} \right) \right] \ .
\end{equation}

Finally, given two disjoint spatial regions $A$ and $B$, the mutual information $I(A,B)$ and Rényi mutual information is given by
\begin{equation}
    I(A, B) = S(A) + S(B) - S(A \cup B) \ ,
\end{equation}
\begin{equation}
    I_{n}(A, B) = S_{n}(A) + S_{n}(B) - S_{n}(A \cup B) \ ,
\end{equation}
where $S(A \cup B)$ is evaluated by restricting the global correlation matrix $C$ to the combined index set of regions $A$ and $B$.
 
All the calculations in the lattice where made with $N = 8000$ sites. Due to the fermion doubling problem \cite{Rothe:1992nt}, we have to divide by 2 the results obtained from the lattice in order to compare with the ones obtained in the continuum. 

\section{Long distance coefficients for the $n$th RMI}\label{app:longDist}

In this Appendix we explicitly show the coefficients appearing in the long distance expansion of the $n$th RMI \eqref{eq:RMI_longDist} for the Dirac field in $d=3$ and $d=4$, and for the Rarita-Schwinger field in $d=4$.

We recall that for the Dirac case we only focus on the $\ell=0$ contribution. This gives $\mu_\ell=1/2$ and $\mu_\ell=1$ for $d=3$ and $d=4$, respectively. Thus, we evaluate
\begin{align}
\sum_{k=-\frac{n-1}{2}}^{\frac{n-1}{2}}\alpha_{k/n}\big|_{\mu_\ell=1/2}&=\sum_{k=-\frac{n-1}{2}}^{\frac{n-1}{2}}\frac{\of{n^2-4 k^2}^2 \tan^2(k \pi/n)}{64 n^4}\,,\\
\sum_{k=-\frac{n-1}{2}}^{\frac{n-1}{2}}\tilde\beta_{k/n}\big|_{\mu_\ell=1/2}&=\sum_{k=-\frac{n-1}{2}}^{\frac{n-1}{2}}\frac{\of{n^2-4 k^2}^2\of{4 k^2 + 9 n^2}\tan^2(k \pi/n)}{576 n^6}\,,
\end{align}
\begin{align}
\sum_{k=-\frac{n-1}{2}}^{\frac{n-1}{2}}\alpha_{k/n}\big|_{\mu_\ell=1}&=\frac{-155 - 147 n^2 - 105 n^4 + 407 n^6}{241920 n^5}\,,\\
\sum_{k=-\frac{n-1}{2}}^{\frac{n-1}{2}}\tilde\beta_{k/n}\big|_{\mu_\ell=1}&=\frac{2667 - 8060 n^2 - 8526 n^4 - 6300 n^6 + 20219 n^8}{7741440 n^7}\,.
\end{align}
On the other hand, for the Rarita-Schwinger field in $d=4$ the lowest contribution comes from the $\ell=1$ sector, thus $\mu_\ell=2$. So we compute
\small
\begin{align}
    \sum_{k=-\frac{n-1}{2}}^{\frac{n-1}{2}}\alpha_{k/n}\big|_{\mu_\ell=2}&=\frac{(1- n^2) (10731 + 40068 n^2 + 80306 n^4 + 109412 n^6 + 
   128123 n^8)}{2043740160 n^9}\,,\\
    \sum_{k=-\frac{n-1}{2}}^{\frac{n-1}{2}}\tilde\beta_{k/n}\big|_{\mu_\ell=2}&=\frac{(1 - n^2)}{3188234649600 n^{11}}\notag\\
    &\qquad\times(-15559247 + 5366203 n^2 + 116348074 n^4 + 
     291451574 n^6\notag \\
     &\qquad\qquad+ 422181173 n^8 + 
     507316223 n^{10})\,.
\end{align}
\normalsize
Finally, replacing in \eqref{eq:RMI_longDist} we get
\be\label{eq:RMI_Dirac3} 
\begin{split}
    I_n^{\mathrm{Dirac}\,(3)}(\eta)&=\frac{2}{n-1}\Bigg[\of{\sum_k\frac{\of{n^2-4 k^2}^2 \tan^2(k \pi/n)}{64 n^4}}\eta^2\\
    &\quad+\of{\sum_k\frac{\of{n^2-4 k^2}^2\of{4 k^2 + 9 n^2}\tan^2(k \pi/n)}{576 n^6}}\eta^3\Bigg]+\mc O\of{\eta^4}\,,
\end{split}
\ee
\be\label{eq:RMI_Dirac4}  
\begin{split}
    I_n^{\mathrm{Dirac}\,(4)}(\eta)&=\frac{(1 + n) (155 + 302 n^2 + 407 n^4)}{60480 n^5}\,\eta^3\\
    &\quad+\frac{(1 + n) (-2667 + 5393 n^2 + 13919 n^4 + 20219 n^6)}{1935360 n^7}\,\eta^4+\mc O\of{\eta^5}\,,
\end{split}
\ee
and
\small
\be
\begin{split}\label{eq:RMI_RS} 
    I_n^{\mathrm{RS}\,(4)}(\eta)&=\frac{(1 + n) (10731 + 40068 n^2 + 80306 n^4 + 109412 n^6 + 
   128123 n^8)}{255467520 n^9}\,\eta^5\\
    &+(-15559247 + 5366203 n^2 + 116348074 n^4 + 291451574 n^6 + 
     422181173 n^8 + 507316223 n^{10})\\&\qquad\times\frac{(1 + n)}{398529331200 n^{11}}\,\eta^6+\mc O\of{\eta^7}\,.
\end{split}
\ee
\normalsize
Note that, as stated in the main text, the expressions \eqref{eq:RMI_Dirac4} and \eqref{eq:RMI_RS} allow a direct evaluation at $n\to1$. The corresponding results appear in \eqref{eq:I_4Dirac} and \eqref{eq:I_RS}, respectively.

\bibliography{EE}{}
\bibliographystyle{utphys}

\end{document}